\documentstyle[psfig, subfigure, lscape]{mn}

\def\lsimeq{{_<\atop^{\sim}}}
\def\gsimeq{{_>\atop^{\sim}}}
\title{On the nature of the ISO-selected sources in the ELAIS S2 region}
\author[F. Pozzi et al.]
       {F. Pozzi,$^{1,2}$\thanks{e-mail: pozzi@bo.astro.it} P. Ciliegi,$^{2}$ C. Gruppioni,$^{2,3}$
       C. Lari,$^{4}$ P. H\'eraudeau$^{5,6}$, M. Mignoli,$^{2}$ \
  \newauthor   G. Zamorani,$^{2,4}$ E. Calabrese,$^{2}$ S. Oliver$^{7}$ and M.Rowan-Robinson$^{8}$   \\  
       $^1$ Dipartimento di Astronomia, Universit\'a di Bologna, via Ranzani 1,
       I--40127 Bologna, Italy\\  
       $^2$ INAF/Osservatorio Astronomico di Bologna, via Ranzani 1, I--40127 Bologna, Italy\\
       $^3$ INAF/Osservatorio Astronomico di Padova, vicolo dell'Osservatorio 5,
I-35122 Padova, Italy\\
       $^4$ Istituto di Radioastronomia di Bologna, via Gobetti 101, I--40129 Bologna,
Italy\\
       $^5$ Max-Plank-Institut f\"{u}r Astronomie, K\"{o}nigstuhl 17, D-69117,
Heidelburg, Germany\\
       $^6$ Kapteyn Astronomical Institute, Landleven 12, 9747 AD Groningen,
The Netherlands \\
       $^7$ Astronomy Centre, CPES, University of Sussex, Falmer, Brighton
BN19QJ \\
       $^8$ Imperial College of Science, Technology and Medicine, Prince
Consort Road, London SW72BZ \\} 

\begin{document}
\maketitle

\begin{abstract}

We have studied the optical, near-IR and radio properties of a complete sample
of 43 sources detected at 15 $\mu$m in one of the deeper ELAIS repeatedly
observed region. The extragalactic objects in this sample have 15-$\mu$m flux
densities in the range 0.4$-$10 mJy, where the source counts start
diverging from no evolution models. About 90 \% of the sources (39 out of 43)
have optical counterparts brighter than I=21 mag. Eight of these 39 sources
  have been identified with stars on the basis of imaging data, while for
  another 22 sources we have obtained optical spectroscopy, reaching a high
identification percentage (30/43, $\sim$70 \%). All but one of the 28 sources
with flux density $>$ 0.7 mJy are identified. Most of the extragalactic objects are normal spiral or starburst galaxies at
moderate redshift ($z_{med}\sim$0.2); four objects are Active Galactic Nuclei. We have used the 15-$\mu$m, H$\alpha$ and 1.4-GHz luminosities as indicators of star-formation rate and we have compared the results obtained in these three bands.
While 1.4-GHz and 15-$\mu$m estimates are in good agreement, showing that our galaxies are
forming stars at a median rate of $\sim$40 M$_{\odot}$yr$^{-1}$, the raw
H$\alpha$-based estimates are a factor $\sim$5-10 lower and need a mean correction of $\sim$2 mag to be brought on the same scale as the other two indicators. A
correction of $\sim$2 mag is consistent with what suggested by the Balmer decrements H$\alpha$/H$\beta$ and by the optical
colours. Moreover, it is intermediate between the correction found locally for normal
spirals and the correction needed for high-luminosity 15-$\mu$m objects,
suggesting that the average extinction suffered by galaxies increases with infrared luminosity.

\end{abstract}


\section{Introduction}

Deep 15-$\mu$m galaxy source counts performed in recent years with several ISOCAM
surveys have revealed a new population of faint sources, whose observed source
density is in large excess with respect to the Euclidean predictions. This evidence was first highlighted by the 
deep/ultra-deep surveys (0.05$-$4 mJy), where at flux
densities fainter than about 1 mJy the counts show a strong divergence from
no-evolution models, with an increasing difference that reaches a factor of
10 around the faintest limits (0.05$-$0.1 mJy; Elbaz et al. 1999). Recently, the evidence
of strong evolution has been confirmed also by the source counts derived in
the S1 field of the shallower European Large Area Survey (ELAIS; Oliver et al. 2000) through a new and independent data reduction
technique (Gruppioni et al. 2002). These counts, based on a
large number of extragalactic sources ($\sim$350) detected over a wide area (4 sq. deg.),
sample with a high statistical significance the previously poorly covered flux
density range between IRAS and the deep ISOCAM surveys (0.5$-$100 mJy). The ELAIS differential counts show a significant change in slope occurring
around 2 mJy, from $\alpha\sim$2.35 at brighter fluxes up to $\alpha\sim$3.6 for fainter
fluxes and down to the survey limits. This is in qualitative agreement with
previous results, although the ELAIS counts are somewhat steeper and lower
than the others at faint fluxes (see Gruppioni et al. 2002). Different authors have interpreted and modeled the observed
15-$\mu$m source counts (i.e. Xu 2000; Rowan-Robinson 2001; Chary \& Elbaz 2001;
Franceschini et al. 2001). Common to all the models is the assumption of
strong evolution for dusty star-formation in starburst galaxies. 

In this context, the key instrument to put observational constraints to the proposed models and to
study directly the nature of the sources responsible for the observed strong evolution,
is photometric and spectroscopic identification of
sources at different 15-$\mu$m flux density levels. At faint fluxes, detailed identification studies have been published for three deep
fields: the HDF-N, the HDF-S and the CFRS 1415+52 field. The three
  surveys reach the following limiting sensitivities at 15-$\mu$m: 220 $\mu$Jy for the HDF-N and
  HDF-S (with a completeness of $\sim$90 \% and $\sim$30-40 \% respectively,
  see Aussel et al. 1999 and Oliver et al. 2002)
  and 350 $\mu$Jy for the CFRS 1415+52 (with a completeness of $\sim$100 \% at
  this flux level, see Flores et al. 1999). Although now the HDF-N is $>$95 \% spectroscopically complete
(Franceschini et al. 2001; Elbaz et al. 2002), the only published data
(Aussel et al. 1999) are relative to a subsample of 26 identifications out of
  a total 15-$\mu$m sample of 46 sources. In the
shallower HDF-S survey, 13 out of 26
sources (50\%) have a spectroscopic redshift (Mann et al. 2002), while in the CFRS 1415+52 survey $\sim$40\%
(17/41) of the sources have a spectroscopic redshift (Flores et
al. 1999). The median redshifts found are $\sim$0.6, $\sim$0.5
and $\sim$0.76 for the HDF-N, HDF-S and CFRS 1415+52 respectively (the peak
of the HDF-N distribution shifts towards z$\sim$0.7 considering the redshift
distribution of the larger sample discussed in Franceschini et al. 2001). The majority of the sources detected
in the CFRS and HDF-S fields show `e(a)' optical spectra 
(see Poggianti et al. 1999), typical of either post-starburst systems or
active starburst galaxies obscured by dust. While Flores et al. (1999) support
the first hypothesis, Rigopoulou et al. (2000), studying a subsample of high
redshift HDF-S galaxies, favour the second hypothesis
and conclude that the HDF-S sources are very powerful
starburst galaxies hidden by large amount of dust.

In this paper, we will present the first results of the identifications of 15-$\mu$m sources at intermediate flux density levels in the repeated S2 field
of the ELAIS survey (Oliver et al. 2000). The S2 field is one of the smaller
areas of the survey and data in this field have been reduced with the {\it
  LARI technique}, which was used also for the reduction of the main S1 Southern field (Lari et al. 2001,
hereafter L01). Given its larger area and shallower depth with respect to
  the deeper fields, S2 is very well suited to provide useful information about the sources
in an interesting flux density range (0.4$-$10 mJy). At these fluxes there is
evidence of strong evolution in the starburst galaxy population. 

The layout of the paper is as follows. In section 2 we describe the 15-$\mu$m
sample; in section 3 we present the multiwavelength follow-up and the optical
photometric identifications; in section 4 we present the spectroscopic results, including
spectral classification for the optical counterparts; in section 5 we discuss
the global properties of the 15-$\mu$m sources. In the last section we present
our conclusions. We adopt $H_0=75$ km s$^{-1}$ Mpc$^{-1}$, $\Omega_{m}$=0.3
and $\Omega_{\lambda}$=0.7 throughout the paper.

\begin{figure*}
\centerline{
\psfig{figure=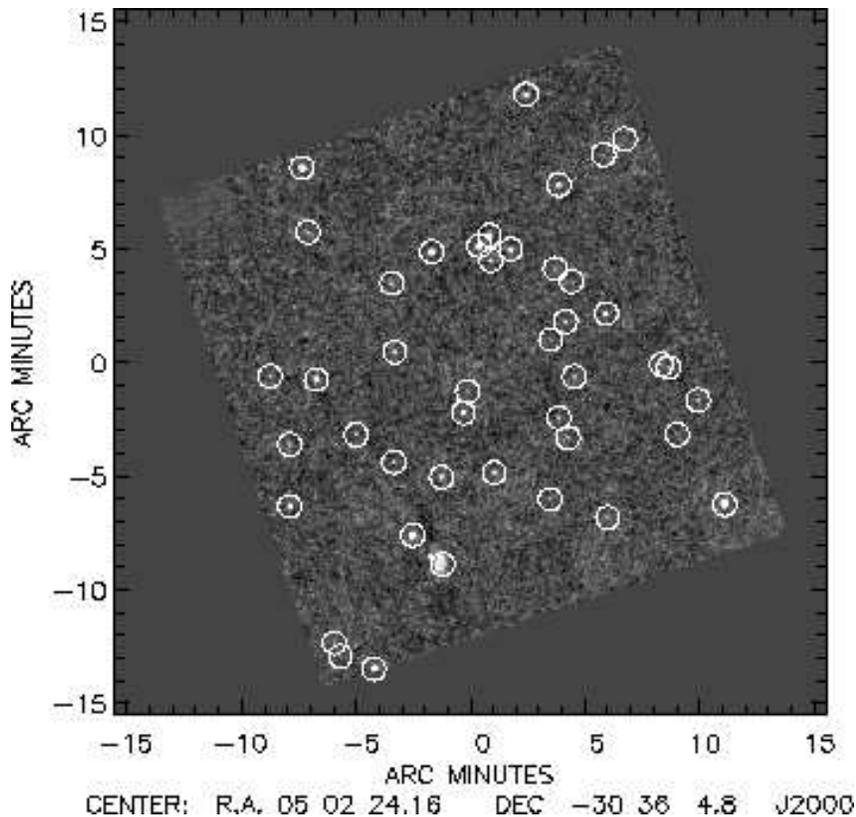,width=11.2cm}}
\caption{Grey scale image of the S2 ELAIS field ($\sim21\times21$ square
  arcminutes). The circles indicate the location of all our detected sources
  (43 objects).}
\label{s2_map}
\end{figure*}

\section{The MIR sample}

The ELAIS survey (Oliver et al. 2000) is the largest survey performed with the Infrared Space Observatory (ISO) at 6.7 $\mu$m, 15 $\mu$m,
90 $\mu$m and 175 $\mu$m.  In particular, the 15-$\mu$m survey (performed with the ISOCAM instrument) covers an
area of  $\sim$12 deg$^2$, divided in 4 main fields and several smaller
areas. The S2 field is one of the smaller areas: centered at $\alpha$(2000) = 05$^h$ 02$^m$ 24.5$^s$, $\delta$(2000) = $-30^{\circ}$
36$^{\prime}$ 00$^{\prime \prime}$, it consists of a single raster of
$\sim21^{'}\times21^{'}$ and covers $\sim$441 arcmin$^2$. Since it has been observed four times, it
is approximately a factor of two deeper than the main survey.  

The 15-$\mu$m data have been reduced and analysed using the {\it LARI
technique}, whose application to the main field S1 has been described in
detail in L01. Since S2 is a repeated field, the procedure of reduction has been the same as that used for
the reduction of the central repeated part of S1 (S1$\_$5), though the
combined rasters in S2 are 4 instead of 3. However, we have used an upgraded version of the software used for S1 improving the mapping procedure
and the checks on individual sources (Lari et al. 2002; Vaccari et al. in
preparation). 

First, the algorithm of reduction has been applied to each of the 4
independent observations separately. For each observation, two maps have been
created, one for the signal and one for the noise. Sources have been
extracted separately in each field and a first cross-correlation with the
optical CCD catalogue in the I band has been computed in order to put the four
15-$\mu$m observations on the same optical astrometric reference system. The 4 different observations have then been combined and
the sources extracted in the combined map (above the 5$\sigma$ threshold). Finally, the
source list has been checked and the total fluxes computed through the procedure
of `autosimulation' described in L01. 

In Figure \ref{s2_map} the final map obtained for S2 is shown (typical $rms$ is  $\sim$13 $\mu$Jy pixel$^{-1}$, somewhat better than the $rms$ of  $\sim$16 $\mu$Jy pixel$^{-1}$, obtained in the S1\_5 repeated
field).
Since the reduction method has been largely tested in the S1 field, in this work we have simulated only 40
sources of 1 mJy, to give an estimate of the completeness, of the flux scale and of
the flux density and positions uncertainties. Almost all the injected
sources are detected (37/40) at $\ge5\sigma$. Thus, our sample is nearly complete
($\sim$93\%) at 1 mJy. At 0.7 mJy, we expect our sample to be $>$ 30\% complete. This was the 
   completeness level found for the S1$\_$5 field at the same flux level
   (repeated 3 times instead of 4). The systematic
flux bias found is 0.88$\pm$0.04 (i.e. fluxes must
be divided by 0.88, see L01); the fluxes were corrected by this factor. The flux errors have been
computed applying the relations found for S1$\_$5 (see eq. 3 of L01), by considering for each S2 source its $rms$, $S/N$ and $f_{0}$
(`theoretical' peak flux, see L01). 
The positional errors for
each source is the combination of the uncertainties due to the reduction method ($\sigma_{s+r}$) and the uncertainties in
the pointing accuracy ($\sigma_{p}$) (see eq. 6,7 in L01). While for the reduction contribution we have assumed the expression
found for S1\_5, we have computed actual values for the pointing accuracy in
S2 by cross-correlating the ISO sources with the I
band optical catalogue (see Section \ref{photometric_sec}). Uncertainties in the  pointing accuracy of about 0.2-0.3$^{\prime \prime}$
have been obtained. These errors are negligible with respect to $\sigma_{s+r}$
which is of the
order of 1-2$^{\prime \prime}$, depending on the $S/N$ of each source\footnote{The
flux and positional errors computed following the above procedures must be considered as conservative
estimates of the real uncertainties affecting the S2 data (repeated 4 times), since
the relations found S1$\_$5 (repeated 3 times) have been used.}.
A catalogue of 43 objects, with flux density in the range 0.4$-$10 mJy, has been obtained.
The catalogue is presented in Table \ref{catalogue_tab}. The extragalactic source counts
  drawn from this catalogue are well consistent with those obtained in S1$\_$5
  over the same flux density range (0.7-6 mJy).

\begin{table*}
\begin{minipage}{150mm}
\caption{The 15-$\mu$m Catalogue in the ELAIS Southern Area S2.}
\label{catalogue_tab}
\scriptsize
\begin{tabular}{rcrrcccrcc} \\\hline \hline
N    &Name &  RA      &   DEC         &  $\sigma$(RA) & $\sigma$(DEC) & S$_{peak}$ &{\it S/N}  & S$_{tot}$ & $\sigma$(S$_{tot}$)  \\
     &     &  (J2000) & (J2000)       &  ('')          &  ('')         & (mJy)    &     & (mJy) &  (mJy)  \\ \hline
1   &    ELAISC15$\_$050143-303528&   05 01 43.41  & -30 35 28.18& 1.2 &1.1 &0.080&   6.17&  0.523&  0.087 \\ 
2   &    ELAISC15$\_$050147-303227&   05 01 47.40  & -30 32 27.91& 1.2 &1.1 &0.079&   6.78&  0.484&  0.074 \\ 
3   &    ELAISC15$\_$050147-302944&   05 01 47.47  & -30 29 44.26& 0.8 &0.6 &0.600&  45.75&  3.446&  0.382 \\ 
4   &    ELAISC15$\_$050149-304438&   05 01 49.85  & -30 44 38.81& 0.8 &0.7 &0.684&  37.44&  4.177&  0.466 \\ 
5   &    ELAISC15$\_$050151-304148&   05 01 51.12  & -30 41 48.06& 1.2 &1.1 &0.091&   6.90&  0.602&  0.094 \\ 
6   &    ELAISC15$\_$050152-303519&   05 01 52.83  & -30 35 19.20& 0.8 &0.6 &0.582&  40.82&  3.567&  0.397 \\ 
7   &    ELAISC15$\_$050156-302343&   05 01 56.53  & -30 23 43.00& 1.3 &1.2 &0.076&   5.53&  0.472&  0.081 \\ 
8   &    ELAISC15$\_$050157-302307&   05 01 57.75  & -30 23 07.47& 1.3 &1.2 &0.107&   5.70&  0.712&  0.124 \\ 
9   &    ELAISC15$\_$050200-303253&   05 02 00.93  & -30 32 53.52& 0.9 &1.0 &0.155&  11.77&  0.921&  0.115 \\ 
10  &    ELAISC15$\_$050204-302234&   05 02 04.58  & -30 22 34.81& 0.8 &0.7 &0.831&  30.08&  6.255&  0.712 \\ 
11  &    ELAISC15$\_$050208-303934&   05 02 08.17  & -30 39 34.19& 0.9 &1.0 &0.187&  11.38&  1.094&  0.137 \\ 
12  &    ELAISC15$\_$050208-303141&   05 02 08.62  & -30 31 41.09& 1.0 &1.0 &0.134&   9.67&  0.788&  0.103 \\ 
13  &    ELAISC15$\_$050208-303631&   05 02 08.75  & -30 36 31.45& 1.0 &1.0 &0.153&  10.73&  0.948&  0.122 \\ 
14  &    ELAISC15$\_$050212-302828&   05 02 12.41  & -30 28 28.07& 0.8 &0.6 &0.698&  53.11&  4.610&  0.511 \\ 
15  &    ELAISC15$\_$050216-304056&   05 02 16.31  & -30 40 56.89& 0.8 &0.6 &0.790&  67.43&  4.782&  0.528 \\ 
16  &    ELAISC15$\_$050218-303100&   05 02 18.27  & -30 31 00.55& 0.9 &0.9 &0.180&  13.77&  1.105&  0.134 \\ 
17  &    ELAISC15$\_$050218-302710&   05 02 18.54  & -30 27 10.93& 1.0 &1.0 &0.136&   9.99&  0.815&  0.106 \\ 
18  &    ELAISC15$\_$050222-303351&   05 02 22.74  & -30 33 51.58& 0.8 &0.7 &0.366&  28.10&  1.926&  0.216 \\ 
19  &    ELAISC15$\_$050223-303447&   05 02 23.55  & -30 34 47.08& 1.2 &1.1 &0.082&   6.25&  0.491&  0.077 \\ 
20  &    ELAISC15$\_$050225-304112&   05 02 25.96  & -30 41 12.39& 0.8 &0.6 &0.653&  49.29&  4.040&  0.448 \\ 
21  &    ELAISC15$\_$050228-304140&   05 02 28.05  & -30 41 40.47& 0.8 &0.7 &0.464&  36.45&  2.900&  0.324 \\ 
22  &    ELAISC15$\_$050228-304034&   05 02 28.30  & -30 40 34.24& 0.8 &0.8 &0.319&  23.60&  1.971&  0.225 \\ 
23  &    ELAISC15$\_$050228-303113&   05 02 28.96  & -30 31 13.99& 0.8 &0.9 &0.215&  16.09&  1.358&  0.162 \\ 
24  &    ELAISC15$\_$050232-304103&   05 02 32.37  & -30 41 03.17& 0.8 &0.7 &0.321&  25.03&  1.858&  0.210 \\ 
25  &    ELAISC15$\_$050235-304752&   05 02 35.49  & -30 47 52.37& 0.8 &0.8 &0.348&  19.00&  2.127&  0.247 \\ 
26  &    ELAISC15$\_$050240-303002&   05 02 40.30  & -30 30 02.96& 1.2 &1.1 &0.079&   6.01&  0.564&  0.099 \\ 
27  &    ELAISC15$\_$050240-303704&   05 02 40.48  & -30 37 04.98& 1.3 &1.2 &0.076&   5.05&  0.486&  0.089 \\ 
28  &    ELAISC15$\_$050241-304012&   05 02 41.42  & -30 40 12.24& 1.0 &1.0 &0.125&   9.32&  0.822&  0.113 \\ 
29  &    ELAISC15$\_$050242-303338&   05 02 42.14  & -30 33 38.29& 1.0 &1.0 &0.126&   9.19&  0.773&  0.104 \\ 
30  &    ELAISC15$\_$050242-304354&   05 02 42.19  & -30 43 54.35& 0.8 &0.8 &0.320&  23.82&  2.164&  0.249 \\ 
31  &    ELAISC15$\_$050243-303753&   05 02 43.49  & -30 37 53.32& 1.0 &1.0 &0.119&   9.71&  0.666&  0.086 \\ 
32  &    ELAISC15$\_$050243-303245&   05 02 43.98  & -30 32 45.11& 1.2 &1.1 &0.096&   7.04&  0.620&  0.095 \\ 
33  &    ELAISC15$\_$050244-303938&   05 02 44.64  & -30 39 38.43& 1.3 &1.2 &0.075&   5.56&  0.509&  0.091 \\ 
34  &    ELAISC15$\_$050245-303526&   05 02 45.22  & -30 35 26.80& 1.1 &1.1 &0.101&   7.63&  0.797&  0.130 \\ 
35  &    ELAISC15$\_$050251-304513&   05 02 51.25  & -30 45 13.44& 1.3 &1.2 &0.069&   5.28&  0.447&  0.081 \\ 
36  &    ELAISC15$\_$050251-303813&   05 02 51.73  & -30 38 13.71& 0.8 &0.8 &0.252&  19.35&  1.705&  0.200 \\ 
37  &    ELAISC15$\_$050251-302914&   05 02 51.99  & -30 29 14.25& 1.2 &1.1 &0.078&   6.35&  0.531&  0.088 \\ 
38  &    ELAISC15$\_$050255-304554&   05 02 55.59  & -30 45 54.50& 1.3 &1.2 &0.070&   5.09&  0.394&  0.067 \\ 
39  &    ELAISC15$\_$050302-303559&   05 03 02.96  & -30 35 59.59& 1.3 &1.2 &0.072&   5.21&  0.444&  0.078 \\ 
40  &    ELAISC15$\_$050304-303551&   05 03 04.45  & -30 35 51.09& 1.2 &1.1 &0.093&   6.57&  0.732&  0.129 \\ 
41  &    ELAISC15$\_$050306-303254&   05 03 06.13  & -30 32 54.05& 1.3 &1.2 &0.066&   5.35&  0.432&  0.078 \\ 
42  &    ELAISC15$\_$050310-303424&   05 03 10.46  & -30 34 24.25& 1.1 &1.1 &0.111&   8.33&  0.708&  0.100 \\ 
43  &    ELAISC15$\_$050315-302948&   05 03 15.79  & -30 29 48.89& 0.8 &0.6 &1.603& 115.82& 10.275&  1.132 \\ \hline \hline
\end{tabular}
\footnotesize
Notes: Col.(1): the ISO source number. Col.(2): the source name. Col.(3),(4): right ascension and declination at
equinox J2000. Col.(5),(6): the positional accuracy. Col.(7): the source peak flux (in
mJy pixel$^{-1}$). Col.(8): the detection level (signal-to-noise ratio). Col.(9),(10): the total flux
(corrected for the flux bias) and its error (in mJy).
\end{minipage}
\end{table*}

\begin{table*}
\begin{minipage}{170mm}
\caption{Multi-band properties and spectroscopic redshifts of the 15-$\mu$m
  sources in the ELAIS S2 field.}
\label{followup_table}
\scriptsize
\begin{tabular}{rlccrrrrrrcccrc}  \\\hline \hline
N  &S$_{tot}$  & I& $\Delta_{I}$ & LR         &Rel& U     & B     & R$_{63F}^a$ &K$^\prime$$^{b}$  &z   &$\Delta_{1.4GHz}$$^{c}$& S$_{1.4}^{peak}$&  S$_{1.4}^{tot}$ \\  
    &  (mJy)&         &    ('')     &         &   &       &         &        &       &&('')  &(mJy/beam) & (mJy) \\ \hline
    1 &  0.52  &20.54  &   2.08  &       3.89  &   0.97  & $>$21.00&$>$24.50 &$>$21.00&17.15  & &1.59 & 0.11      &  0.09   \\
    2 &  0.48  &17.70  &   1.87  &      45.04  &   0.99  &  19.72  &  19.66  &  18.29 &15.56  & &     &$<$0.08    &      \\
    3 &  3.45  &17.18  &   0.87  &     183.72  &   0.99  &  19.21  &  18.86  &  17.46 &15.03  &0.127 &0.95 & 0.48      &  0.38  \\
    4 &  4.18  &14.15  &   0.77  &     241.89  &   0.99  &  16.52  &  16.07  &  $-$  &12.28  &0.050 &1.59 & 0.43      &  0.43  \\
    5 &  0.60  &19.88  &   1.06  &      24.09  &   0.99  &$>$21.00 &  22.86  &  20.80 &16.98  & &5.2  & 0.09      &  0.09  \\
    6 &  3.57  &17.58  &   1.14  &     121.35  &   0.99  &  18.77  &  18.76  &  17.86 &16.06  &1.813 &     & $<$0.08   &       \\
    6 &        &17.29  &   2.95  &       0.96  &   0.00  &         &         &        &       & &     &            &      \\
    7 &  0.47  &19.66  &   2.85  &       2.96  &   0.75  &$>$21.00 &$>$24.50 &$>$21.00&$n.a.$ & &     & $<$0.08   &      \\        
    7 &        &15.71  &   3.98  &       0.84  &   0.21  &         &         &        &       & &     &            &       \\  

    8 &  0.71  &19.46  &   1.47  &      18.42  &   0.99  &  20.63  &  20.84  &  19.91 &$n.a.$ &0.600 &     & $<$0.08   &      \\
    9 &  0.92  &17.83  &   1.72  &      51.55  &   0.99  &  20.18  &  20.06  &  18.40 &15.41  &0.308 &0.84 & 0.21      &  0.26  \\
   10 &  6.26  &10.54  &   0.92  &    4828.14  &   0.99  &  11.80  &  10.92  &   8.71 &$n.a.$ &0 &     & $<$0.08   &        \\
   11 &  1.09  &20.64  &   1.06  &      12.75  &    0.99 &$>$21.00 &  23.77  &$>$21.00&17.75  &0.627 &        & $<$0.08   &     \\
   12 &  0.79  &17.16  &   3.83  &       0.53  &   0.84  &  18.54  &  18.69  &  17.40 &15.19  & &     & $<$0.08   &       \\
   13 &  0.95  &19.25  &   1.51  &      19.15  &   0.99  &$>$21.00 &  21.65  &  20.01 &16.63  &0.450 &2.60 & 0.12      &  0.08  \\
   14 &  4.61  &17.46  &   1.15  &     126.26  &   0.99  &  17.44  &  18.00  &  17.25 &15.38  &0.862 &4.11 & 0.09      &  0.09  \\
   15 &  4.78  &11.23  &   0.76  &    5128.13  &   0.99  &  14.45  &  13.48  &   9.98 & 9.82  &0 &     & $<$0.08   &       \\
   16 &  1.11  &18.60  &   1.31  &      45.65  &   0.99  &  20.90  &  20.39  &  19.56 &15.94  &0.128 &2.01 & 0.14      &  0.14  \\
   17 &  5.6   &12.52  &   0.01  &     554.57  &   0.99  &  15.07  &  14.00  &  $-$   &10.51  &0.020 &8.50$^\star$ & 0.13      &  0.18  \\
   18 &  1.93  &17.05  &   0.74  &     201.17  &   0.99  &  19.89  &  19.28  &  17.42 &14.64  &0.111 &0.61 & 0.40      &  0.33  \\
   19 &  0.49  &$>$22.00&        &             &         &$>$21.00 & $>$24.50&$>$21.00&$>$18.75&&      & $<$0.08   &       \\
   20 &  4.04  &17.56  &   0.07  &     289.72  &   0.99  &  20.72  &  20.04  &  18.17 &14.91  &0.191 &2.17 & 0.69      &  0.87  \\
   21 &  2.90  &16.86  &   0.63  &     218.17  &   0.99  &  19.40  &  19.18  &  17.43 &14.56  &0.191 &0.87 & 0.33      &  0.45  \\
   22 &  1.97  &11.13  &   0.99  &    3715.93  &   0.99  &  12.14  &  12.40  &   9.17 & 9.91  &0 &     & $<$0.08   &       \\
   23 &  1.36  &18.09  &   0.50  &     154.87  &   0.99  &  20.92  &  20.26  &  18.92 &15.73  &0.138 &1.67 & 0.10      &  0.11  \\
   24 &  1.86  &17.84  &   1.87  &      32.13  &   0.99  &  20.84  &  20.09  &  18.51 &15.62  &0.123 &2.40 & 0.14      &  0.12  \\
   25 &  2.13  &14.76  &   1.94  &      30.91  &   0.99  &   $n.a.$&  16.10  &  $-$   &$n.a.$ &0.016 &2.51 & 0.14      &  0.14  \\
   26 &  0.55  &$>$22.00&        &             &         &$>$21.00 & $>$24.50&$>$21.00&$>$18.75&&      & $<$0.08   &       \\
   27 &  0.49  &18.25  &   4.00  &       1.02  &   0.91  &$>$21.00 & 21.05   &  19.19 &15.64  & &     & $<$0.08   &       \\
   28 &  0.82  &11.68  &   0.85  &    1749.25  &   0.99  &  12.41  &  12.80  &  $-$   &10.46  &0 &     & $<$0.08   &       \\
   29 &  0.77  &17.12  &   0.89  &     128.63  &   0.99  &  19.56  &  19.08  &  17.31 &15.04  &0.125 &     & $<$0.08   &       \\
   30 &  2.16  &17.71  &   0.61  &     204.91  &   0.99  &  20.81  &  20.05  &  18.29 &15.17  & 0.170&2.11 & 0.41      &  0.42 \\
   31 &  0.67  &12.14  &   1.70  &     437.34  &   0.99  &  14.27  &  13.78  &  11.43 &10.39  & 0&     & $<$0.08   &       \\
   32 &  0.62  &20.39  &   2.49  &       3.27  &   0.97  &$>$21.00 &  23.57  &$>$21.00&17.07  & &2.61 & 0.10      &  0.10  \\
   33 &  0.51  &12.82  &   2.42  &      73.83  &   0.99  &  15.63  &  14.46  &  12.41 &11.18  & 0&     & $<$0.08   &       \\
   34 &  0.80  &19.39  &   2.52  &       4.56  &   0.72  &         &         &        &       &      &         &\\
   34 &        &20.70  &   2.57  &       1.66  &   0.26  &$>$21.00 &$>$24.50 &$>$21.00&17.98  & 0.775& 3.00  &0.33   &0.54           \\
   35 &  0.45  &20.61  &   1.87  &       5.01  &   0.98  &  21.82  &  22.37  &$>$21.00&18.59  & &     & $<$0.08   &       \\
   36 &  1.71  &16.92  &   0.81  &     166.70  &   0.99  &  18.89  &  18.73  &  17.14 &14.81  & 0.150&2.45 & 0.17      &  0.31 \\
   37 &  0.53  &18.65  &   2.71  &       6.82  &   0.67  &  20.20  &  20.19  &  19.41 &16.43  & 0.279&1.68 & 0.16      &  0.13 \\
   37 &        &20.54  &   2.23  &       3.16  &   0.31  &         &         &        &       & &     &            &       \\
   38 &  0.39  &$>$22.00&        &             &         &$>$21.00 & $>$24.50&$>$21.00&$n.a.$ & &     & $<$0.08   &       \\              
   39 &  0.44  &$>$22.00&        &             &         &$>$21.00 & $>$24.50&$>$21.00&$n.a.$ & &     & $<$0.08   &    \\
   40 &  0.73  &11.41  &   0.94  &    2699.30  &   0.99  &  12.70  &  12.68  &  10.32 &$n.a.$ & 0&     & $<$0.08   &       \\
   41 &  0.43  &18.55  &   0.57  &      55.74  &   0.99  &  20.54  &  20.36  &  19.31 &$n.a.$ & &     & $<$0.08   &       \\
   42 &  0.71  &17.89  &   0.69  &     132.31  &   0.99  &  19.38  &  19.38  &  18.47 &$n.a.$ & 0.172&1.75 & 0.10      &  0.16 \\
   43 &  10.28 & 9.83  &   3.21  &       3.86  &    0.97 &  10.9   &  11.10  &    $-$   &$n.a.$ &0&  &    & $<$0.08   &       \\ \hline \hline          
\end{tabular}
\footnotesize
Notes: Col.(1): the ISO source number. Col.(2): the 15-$\mu$m total flux (in mJy). Col.(3): the I band
magnitude of the optical counterpart. Col.(4),(5),(6): the offset (in arcsec)
between the ISO position and the optical counterpart in the I band, its likelihood ratio LR and 
reliability Rel. Col.(7),(8),(9),(10): the magnitude in the U, B, R$_{63F}$
and K$^{\prime}$ bands. The ($-$) symbol in Col.(9) indicates a not reliable
APM magnitude. $n.a.$ in Col.(10) indicates that the K$^\prime$ magnitude is not available. Col.(11): the spectroscopic redshift (see Section \ref{spettro_sec}). Col.(12),(13),(14): the distance (in arcsec) between the ISO and the radio 
counterpart, the 1.4-GHz peak and total fluxes of the radio counterpart. The
($^\star$) symbol for object \#17 indicates a large distance between the ISO
and radio centroids (see Figure \ref{spectra3_fig}).
\end{minipage}
\end{table*}

\section{Multiband Photometric Follow-up} 
\label{photometric_sec}

Optical follow-up has been obtained for the S2 field in the U, B and I bands with the WFI at the ESO 2.2-m Telescope. The optical catalogues are complete down to U $\sim$21.0, 
B $\sim$24.5 and I $\sim$22.0 (Heraudeau et al. in preparation). 
Moreover, a near infrared survey 
in the K$^{\prime}$ band has been obtained (over most of the S2 area) with SOFI at the ESO 
NTT, down to K$^{\prime}$ $\sim$18.75 (Heraudeau et al. in preparation). 
The APM catalogue (Maddox et al. 1990)\footnote{The Automatic Plate Measuring (APM) machine is a
National Astronomy Facility run by the Institute of Astronomy in
Cambridge. See http://www.ast.cam.ac.uk/$\sim$apmcat/.}
in the R band is also available down to R$_{63F}$ $\sim$21.
Finally, the whole S2 area has been surveyed in the radio band at 1.4 GHz 
with the Australia Telescope Compact Array (ATCA) down to a 
5$\sigma$ flux limit of 0.13 mJy (Ciliegi et al. in preparation). 

\subsection{Optical and near infrared identification of the ISO sources} 

We define the I band catalogue as our master optical catalogue, which we used to search for the optical counterparts of the ISO sources using the likelihood 
ratio technique described by Sutherland \& Saunders (1992). The 
likelihood ratio LR is the ratio between the probability that a given source at the observed position and with the measured magnitude is the true optical counterpart, and the probability that the same source is a chance background object.
For each source we adopted an elliptical Gaussian distribution for the
positional errors with the standard deviation in RA and DEC reported in Table
\ref{catalogue_tab} and assuming a value of 0.5 arcsec as the optical position
uncertainty. 

For each optical candidate 
we estimated also the reliability (Rel), by taking into account, when
necessary, the presence 
of other optical candidates for the same ISO source (Sutherland \& 
Saunders 1992). Once the likelihood ratio (LR) has been calculated for all the optical candidates, 
one has to choose the best threshold value for LR (LR$_{th}$) to 
discriminate between spurious and real identifications.  
As the LR threshold we adopted LR$_{th}=$0.5. With this value, all the optical
counterparts of the ISOCAM sources with only one identification (the majority
in our sample) and LR$>$LR$_{th}$ have a reliability greater than $\sim$0.8
(we assumed a value of Q=0.9 for the probability that an optical counterpart
of the ISOCAM source is brighter than the magnitude limit of the optical
catalogue, see Ciliegi et al. 2003 for more details). With this threshold value we find 39 ISO 
sources with a likely identification (four of which have two optical 
candidates with  LR$_{th}>$0.5). The same number of ISO/optical associations
would be found using the less conservative value of LR$_{th}=$0.2 (i.e. we do
not have optical counterparts with 0.2$<$LR$<$0.5).
A summary of the results for the identification of the 43 ISO 
sources in the optical, near infrared and radio bands is given 
in Table \ref{followup_table}.  As shown in column 4 of Table \ref{followup_table}, all the 
likely optical counterparts lie within 4 arcsec from the ISO position, 
the majority of them having an ISO-optical offset smaller than 2 arcsec. 

The reliability (Rel) of each optical identification is always very 
high ($>$0.98 for 90\% of the sources), except for the four cases 
where more than one optical candidate with  LR$_{th}>$0.5 is present for the 
same ISO source. For these four sources we assumed that the object with the 
highest likelihood ratio value is the real counterpart of the ISO source. \\
The number of expected real identifications (obtained adding the reliability
of all the objects with  LR$_{th}>$0.5) is about 38, $i.e.$ we expect that
$\sim$ 1 of the 39 proposed  ISO-optical associations may be 
spurious positional coincidences. 

Starting from the I band optical position of the 39 
proposed  ISO-optical associations we looked for U, B, R$_{63F}$ and  K$^{\prime}$ 
counterparts using a maximum distance of 1 arcsec (1.5 for the R$_{63F}$ filter). For 10 of the 43 
ISO sources  K$^{\prime}$ band data are not available (quoted as $n.a.$ in
column 10 of Table 2). We found 31 ($\sim$72 \%) counterparts in the U band, 
37 ($\sim$86 \%) counterparts in the B band, 32 ($\sim$74 \%) counterparts in the R$_{63F}$ band
and 31 ($\sim$72 \%) counterparts in the 
K$^{\prime}$ band. The same results are obtained using a search radius 
of 2 arcsec. The APM magnitudes of sources \#4, 17, 25, 28 and 43 have not been
reported since they appeared to be not reliable from a comparison with
the magnitudes in the other bands. Three of these sources (\#4, 17 and 25) are the
brightest and more extended galaxies in our sample (see Figure \ref{spectra3_fig} for a $2{\times}2$ arcmin map of source \#17), while sources \# 28 and 43
are stars of $12^{th}$ and $11^{th}$ magnitude respectively.

Finally, for the 4 ISO sources without a likely optical counterpart in the 
I band (sources 19, 26, 38 and 39)  we looked for possible 
counterparts in the U, B, R$_{63F}$ K$^{\prime}$ and radio bands using a maximum 
distance of 5 arcsec from the ISO position. However, no counterparts have been 
found for these 4 sources (see Section
  \ref{multiband_sec} for a discussion relative to these sources).

\begin{figure*}
\vspace{23.cm}
\caption{EFOSC2 spectra of 21 of the 22 spectroscopically identified extragalactic objects with their corresponding I band CCD images (data for source \#17 are shown separately in Figure \ref{spectra3_fig}). The size of each image is 1${\times}1$ arcmin. Contour levels of the 15-$\mu$m emission corresponding to
[3,4,5,6,7,8,9,10,12,15,20,25,30,50,100]
$\sigma$ are superimposed on each optical image. }
\label{spectra_fig}
\end{figure*}

\addtocounter{figure}{-1}
      
\begin{figure*}
\vspace{23.cm}
\caption{continue}
\label{}
\end{figure*}

\addtocounter{figure}{-1}
      
\begin{figure*}
\vspace{23.cm}
\caption{continue}
\label{}
\end{figure*}
       
\addtocounter{figure}{-1}
        
\begin{figure*}
\vspace{10.cm}
\caption{continue}
\label{}
\end{figure*}

\begin{figure*}
\vspace{23cm}
\caption{I band CCD images of the 13 sources with no spectral information and
  not classified as stars. A `B' symbol in the top left hand corner of the
  thumbnails indicates an optically blank object (sources \#19, 26, 38 and 39). The size of each image is 1${\times}1$ arcmin. Contour levels of the 15-$\mu$m emission corresponding to
[3,4,5,6,7,8,9,10,12,15,20,25,30,50,100] $\sigma$ are superimposed on each optical image.}
\label{spectra2_fig}
\end{figure*}

\addtocounter{figure}{-1}
\begin{figure*}
\vspace{5cm}
\caption{continue}
\label{}
\end{figure*}

\begin{figure*}
\vspace{10cm}
\caption{I band CCD images of the 8 sources with no spectral information and
  classified as stars from the photometric data. Sources 10, 15, 22, 28, 40
  and 43 have been found also in the stellar Tycho-2 catalogue (Hog et
  al. 2000). The size of each image is 1${\times}1$ arcmin. Contour levels of
  the 15-$\mu$m emission corresponding to
  [3,4,5,6,7,8,9,10,12,15,20,25,30,50,100] $\sigma$ are superimposed on each optical image.}
\label{spectra4_fig}
\end{figure*} 

\begin{figure}
\centerline{
\psfig{figure=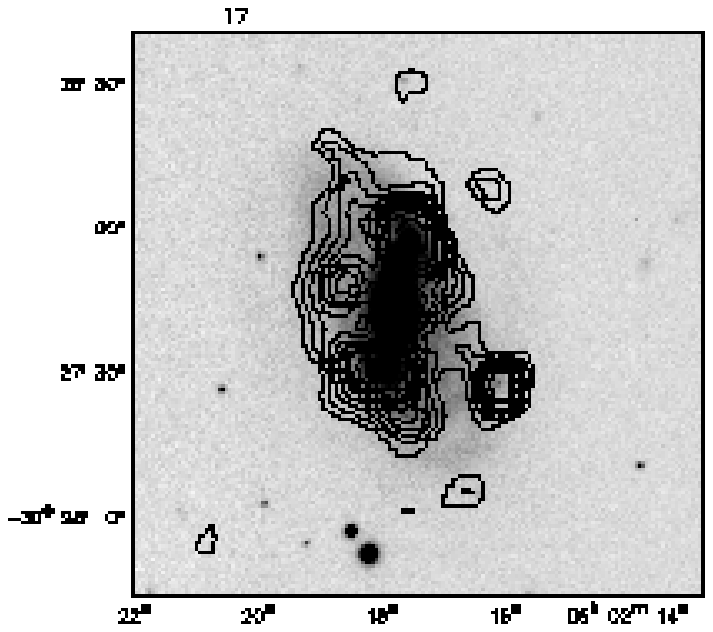,width=5cm}}
\centerline{
\psfig{figure=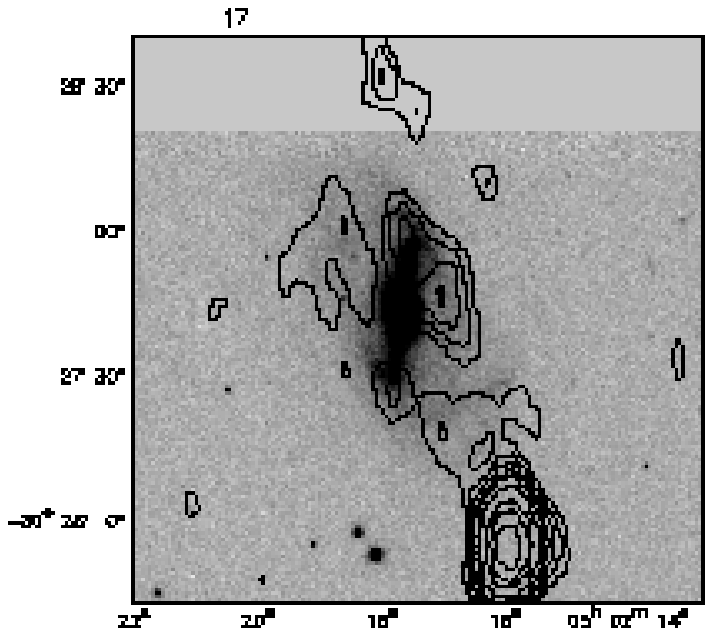,width=5cm}}
\centerline{
\psfig{figure=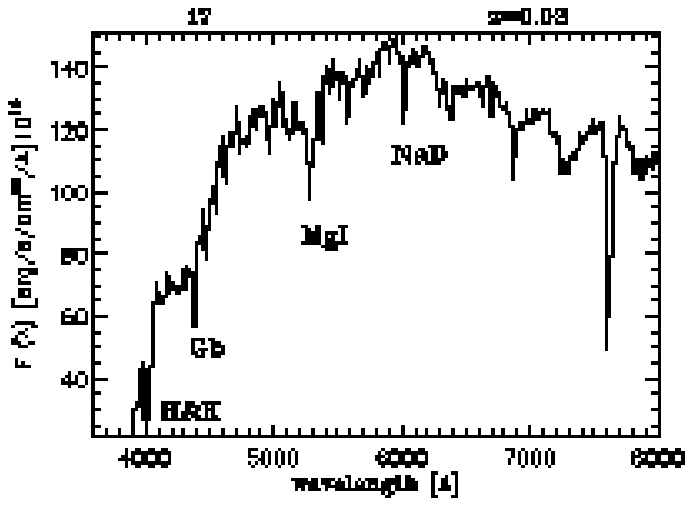,width=5.2cm}}
\caption{{\it Top panel}: Optical I band image of the source \#17 with
    superimposed the contour levels of the 15-$\mu$m emission corresponding to
[3,4,5,6,7,8,9,10,12,15,20,25,30,50,100]$\sigma$. The size of the image is
    2${\times}2$ arcmin. {\it Central panel}: As in the {\it Top panel} with the contour levels of the
    radio emission ([2,3,4,5,6,8,12,15,20,30,50,75,100]$\sigma$) superimposed
    to the K$^{\prime}$ image. {\it Botton panel}: EFOSC2 spectrum of source. }
\label{spectra3_fig}
\end{figure}

\subsection{ISO radio associations}

The radio catalogue in S2 consists of 75 1.4-GHz sources brighter than 
0.13 mJy (5$\sigma$ level). As a first step, a cross correlation 
was performed between 
the radio catalogue and the 43 ISOCAM sources listed in Table \ref{catalogue_tab}. 
We find 13 reliable radio-ISO associations with a positional difference 
smaller than $\sim$5 arcsec (except for the extended source \# 17 where the centroids positions are at $\sim$8.5 arcsec, see Figure \ref{spectra3_fig}). Then, in correspondence of each ISOCAM 
position, we have searched for detection in the radio map down to a 
3$\sigma$ level ($\sim$0.08 mJy), finding 8 additional radio identifications within 5 arcsec.

Assuming a 1.4 GHz source density of $\sim$1000 
sources per square degree at 0.08 mJy (Bondi et al. 2002), a maximum distance of 5 arcsec between the ISO and radio sources 
corresponds to a random association probability (P=1-e$^{-N(S)\pi d^2}$, 
where N(S) is the density of radio sources with flux greater than $S$ and 
$d$ is the distance between the ISO and the radio position) lower 
than 0.006. We therefore expect that essentially all these radio sources are
physically associated with the ISO sources. 

Finally, we have verified that the positions of the radio and optical
counterparts, associated to the same ISO source, are all consistent with each other.
  
\section{Spectroscopy}

\subsection{Observations}
\label{spettro_sec}

We have obtained spectroscopic data for 22 of the 39 likely optical
counterparts. We did not observe 8 objects with likely optical
counterparts (sources \# 10, 15, 22, 28, 31, 33, 40
and 43) since they are easily classified as stars from the photometric
data (see Figure \ref{spectra4_fig}). All of them show, in fact, a clear stellar
appearance and are characterized by very bright I magnitude. Indeed, 6 of them have been found also in the stellar Tycho-2 catalogue (sources \# 10,
15, 22, 28, 40 and 43, Hog et al. 2000). 

The spectroscopic observations have been performed during 1999 December 5-6, using the ESO Faint Object
Spectrograph and Camera Version 2 (EFOSC2) at the 3.6 m ESO Telescope. We used the grism \#6, with spectral range
3860-8070 \AA $~$  and resolution of 4 \AA /pixel (binning 2$\times$2).
The slit used was 1.2-1.5$^{\prime\prime}{\times}5^{\prime}$. The exposure
times varied from a minimum of 120 seconds for the brightest objects
(I $\sim$14 mag) up to a maximum of 2 hours for the faintest objects
(I $\sim$21 mag). To optimize the exposure time, when possible
two objects have been observed simultaneously by rotating the slit. 

\subsection{Data reduction}

The data reduction has been performed with the software IRAF and the add-on package RVSAO, which contains tasks to obtain radial
velocities from spectra using cross-correlation and emission lines fitting techniques. To remove the pre-flash illumination, for each night a median bias (obtained from a sample of ``zero exposures''
taken at the beginning and at the end of the night) has been subtracted from all the frames. To remove the
pixel-to-pixel variations, the frames have been calibrated using flat fields
obtained from an internal quartz halogen lamp located in the dome. To
calculate and subtract the background, a fit has been performed to the intensity
along the spatial direction in the columns adjacent to the target
position. The spectrum for each galaxy was extracted using the
APEXTRACT package. Standard wavelength calibration was carried out using
Helium-Argon lamp exposures, taken at the beginning and the end of each
night. Finally, to calibrate in flux, three standard stars have been observed
each night (LTT 1020, LTT 377 and LTT 3218; Hamuy et al. 1992, 1994). Since several objects have been observed in more than one
exposure (to keep the single integration times shorter than
1200 sec) spectra associated to the same object have been combined together to
get a better S/N ratio.

\subsection{Optical Spectra and classification}

\begin{figure*}
\centerline{
\psfig{figure=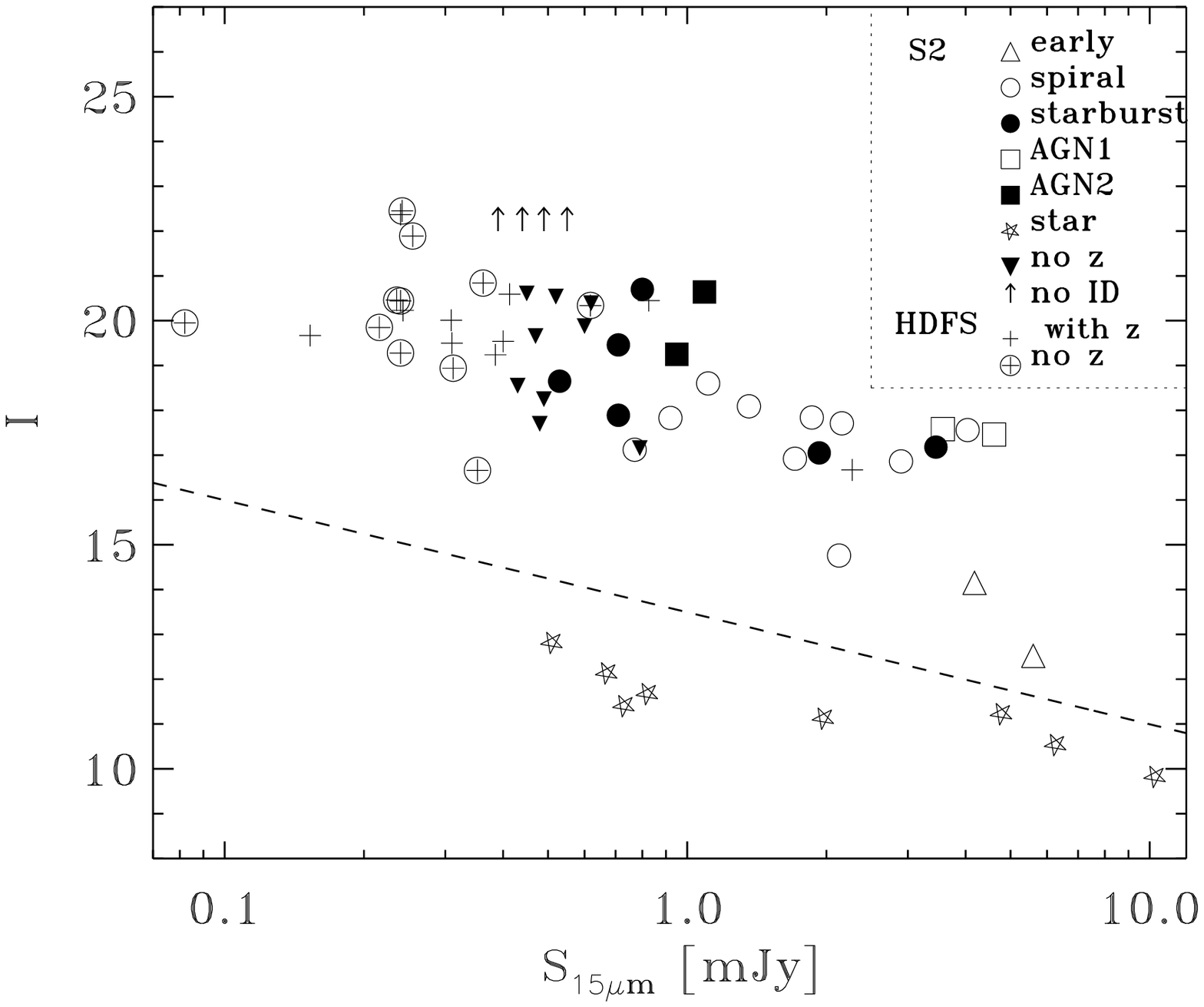,width=8cm}
\psfig{figure=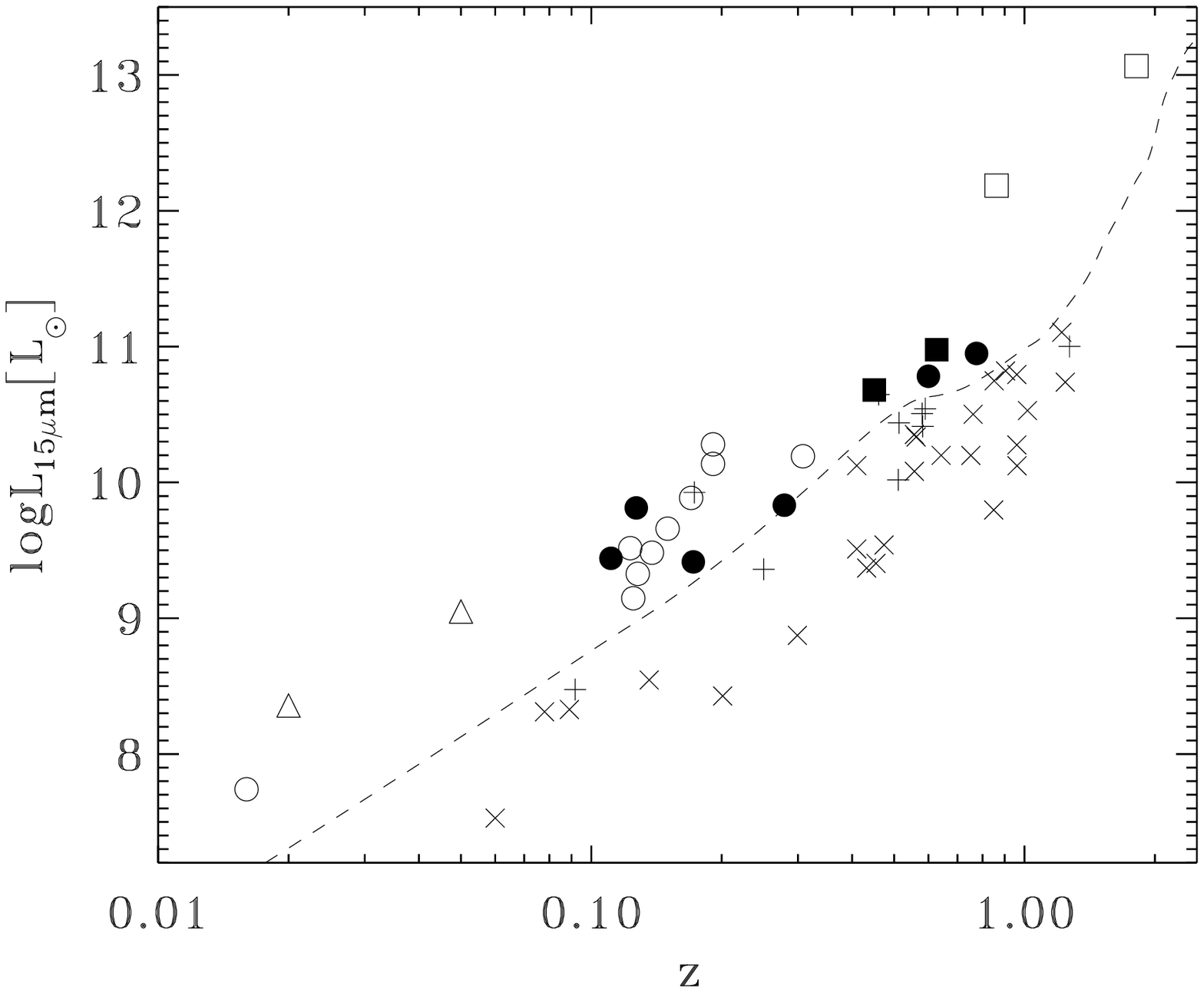,width=8cm}}
\caption{{\it Left panel}: I magnitudes versus 15-$\mu$m flux densities for
  the S2 and HDFS sources. The different symbols represent the different
spectroscopic classes of objects: empty triangles stand for early galaxies, empty circles for normal spiral galaxies, filled circles for starburst galaxies, squares for AGNs (empty for type 1 and filled for type 2) and star symbols
for stars. The filled triangles stand for the objects not observed
spectroscopically, while the vertical arrows are for sources with no optical counterparts. A line of
constant ratio between 15-$\mu$m and I band flux which separates stars from
extragalactic objects is also reported. As a comparison, the data from the HDF-S
are included: the crosses represent the objects spectroscopically identified, while the crosses inside circles represent the objects not observed
spectroscopically (Mann et al. 2002). {\it Right panel}: 15-$\mu$m rest-frame
luminosity (in solar units) versus redshift. Symbols are the same as for the
{\it Left} panel. In this figure also the data from the HDF-N are included
(diagonal crosses, Aussel et al. 1999). The dashed line corresponds to the
  minimum detectable luminosities at different $z$ in S2, considering a limiting
  flux of 0.5 mJy and the 15-$\mu$m $K$-correction of M82.} 
\label{mir_ir_fig}
\end{figure*}

From our spectroscopic observations and reduction we were able to obtain a
reliable redshift determination for all the observed sources (22
objects), reaching a high
identification percentage (30/43, $\sim$70 \%), considering also the eight
stars. Moreover, all but one of the 28 sources
with flux density $>$ 0.7 mJy are identified (see Figure \ref{mir_ir_fig}). In
Figure \ref{spectra_fig}, 21 spectra are presented, together with the
corresponding I band images with superimposed the contour levels of the 15-$\mu$m emission. Source \#17 is presented separately in Figure
\ref{spectra3_fig}, because of its extension. The optical images shown in
Figure \ref{spectra2_fig} correspond to the non-stellar sources
(13 objects) without spectral information; Figure \ref{spectra4_fig} shows the optical images of the stellar sources (8 objects).

Redshifts have been determined by Gaussian-fitting of the emission lines
and via cross-correlation with template spectra for the absorption-line cases. As templates for the cross-correlation we used those of Kinney et al. (1996).
The results of the analysis are presented in Table \ref{spectra_tab}. The line
equivalent width ($EW$) and the $H\alpha$ fluxes reported in Table
\ref{spectra_tab} have been measured using the package SPLOT within the IRAF environment,
comparing the results found with a Gaussian fitting and interactively choosing
the endpoints. Repeated measurements show that the typical uncertainty in the
$EW$ is a few percent for the strongest lines, but can be as high as $\sim$30-40\% for the weakest lines. We were able to separate the $H\alpha$ line from the $[NII]$ line for the majority of our sources. When this was not possible, the $EW$ and flux of $H\alpha+[NII]$
has been measured, estimating the $EW$ and flux of $H\alpha$ by assuming an average ratio $[NII]$/$H\alpha$
$\approx0.5$ as found by Kennicutt (1992). The assumed value of 0.5 is
consistent with our data; in fact from the 6 sources with deblended $H\alpha$ and $[NII]$ we obtain 0.47$\pm$0.03 (these cases have been
highlighted in Table \ref{spectra_tab}). 

 We have classified the objects as galaxies, AGN (type 1 and 2) and stars
 (last column of Table \ref{spectra_tab}). We have first tried to subdivide the galaxy class into different categories according to Poggianti \& Wu (2000, hereafter PW00). This classification is mainly based on two lines ($[OII]$
in emission and $H\delta$ in absorption) and is the more appropriate to
investigate the star-formation properties of galaxies since these two lines
are good indicators of current and recent star-formation
episodes respectively. However, given the limited resolution of our spectra
(especially critical for the weakest lines such as $H\delta$), in the end we used a coarser classification and divided our galaxies into three categories:

\begin{itemize} 
\item{early-type galaxies - elliptical-like spectrum with little ongoing or recent star-formation;}
\item{normal spiral galaxies - with $H\alpha$ or $[OII]$ present ($EW([OII])\le40$); this category includes both the $e(a)$ and the $e(c)$ galaxies of PW00;} 
\item{starburst galaxies - with very strong emission lines ($EW([OII])\ge40$); this category corresponds to the $e(b)$ class of PW00.} 
\end{itemize}

\begin{table*} 
\begin{minipage}{150mm}
\caption{Spectroscopic results and $H\alpha$, 15-$\mu$m and 1.4-GHz rest-frame
  luminosities of the 15-$\mu$m sources in the S2 field.}
\label{spectra_tab}
\scriptsize
\begin{tabular}{cccccccrrr} \\\hline \hline
  N &    z     &EW([OII])& EW(H$\delta$)& EW(H$\alpha$)$^{a}$ &S(H$\alpha$) &L(H$\alpha)$$^{b}$&L(15${\mu}$m)& L(1.4GHz)&Class \\
    &          & (\AA)  &  (\AA) & (\AA)      & (erg cm$^{-2}$s$^{-1}$)10$^{16}$&L$_{\odot}$&L$_{\odot}$& L$_{\odot}$ &\\\hline     
 1  &          &        &        &            &               &       &       &         &  \\
 2  &          &        &        &            &               &       &       &         &  \\
 3  &   0.127  &  42  &  $<$4  &      73*     & 34.7          &   8.12&   9.81&     4.69& starburst\\
 4  &   0.050  &        &  $<$4  &            &               &       &   9.05&     3.90& early-type \\
 5  &          &        &        &            &               &       &       &         &   \\
 6  &   1.813  &        &        &            &               &       &  13.06&         & AGN\_1 \\
 7  &          &        &        &            &               &       &       &         & \\
 8  &   0.600  &  65  & $<$4   &  out       &               &       &10.78  &         & starburst     \\
 9  &   0.308  &   7  & 4-5    & out        &               &       &10.19  &    5.37 & spiral\\
10  &    0     &        &        &            &               &       &       &         & star\\
11  &   0.628  &  49  & $<$4   &  out       &               &       &10.98  &         & AGN\_2      \\
12  &          &        &        &            &               &       &       &         &      \\
13  &   0.450  &   6  &  $<$4  &   out      &               &       &10.68  &    5.23 &AGN\_2 \\
14  &   0.862  &        &        &            &               &       &12.19   &    5.91&AGN\_1 \\
15  &    0     &        &        &           &               &       &       &         &star         \\
16  &   0.127  &  36  &$<$4    &  47      & 5.0           &   6.90&   9.33&     4.27&spiral \\
17  &   0.020  &  out   & $<$4   &            &               &       &   8.36&     2.71&early-type \\
18  &   0.111  &  43  & $<$4   &  37      & 18.3          &7.50   &   9.44&     4.51&starburst\\
19  &          &        &        &            &               &       &       &         & \\
20  &   0.191  &  32  &  $<$4  & 63       & 8.7           &7.96   &  10.28&     5.43&spiral\\
21  &   0.191  &  12  &  5     &      32* & 20.0          &7.99   &  10.14&     5.15&spiral\\
22  &    0     &        &        &            &               &       &       &         & star\\
23  &   0.139  &  22  & 4-5    & 51*      & 17.9          &7.44   &   9.48&     4.23&spiral\\
24  &   0.123  &  25  & $<$4   &  36      & 4.2           &6.91   &   9.52&     4.16&spiral\\
25  &   0.016  &  out   & $<$4   &   15     & 28.8          &5.97   &   7.74&     2.41&spiral\\
26  &          &        &        &            &               &       &       &         &      \\
27  &          &        &        &            &               &       &       &         &      \\
28  &    0     &        &        &            &               &       &       &         &star        \\
29  &   0.125  &   4  &  $<$4  & 11*      &6.7            &7.31   &   9.15&         &spiral   \\
30  &   0.170   &  19 &  $<$4  & 57       &13.0           &7.73   &   9.89&     5.01&spiral\\
31  &    0     &        &        &            &               &       &       &         & star        \\
32  &          &        &        &            &               &       &       &         &       \\
33  &    0     &        &        &            &               &       &       &         & star      \\
34  &   0.775  &  47  &  $<$4  &  out       &               &       &  10.95&     6.60& starburst \\
35  &          &        &        &            &               &       &       &         & \\
36  &   0.150  &  11  & 5      & 45*      & 20.2          &7.99   &   9.66&     4.76& spiral \\
37  &   0.279  &  45  & $<$4   & out        &               &       &9.83   &     4.97& starburst \\
38  &          &        &        &            &               &       &       &         &      \\
39  &          &        &        &            &               &       &       &         &      \\
40  &    0     &        &        &            &               &       &       &         &star       \\
41  &          &        &        &            &               &       &       &         &      \\
42  &   0.173  &  45  & $<$4   &  79*     &21.8           & 7.93  &   9.41&     4.60&starburst\\
43  &    0     &        &        &            &               &       &       &         & star     \\\hline \hline
\end{tabular} 
\footnotesize
Notes: Col.(1): the ISO source number. Col.(2): the measured spectroscopic
  redshift. Col.(3),(4),(5): the equivalent widths at rest of
$[OII](\lambda=3727$\AA) $~$ and $H\alpha(\lambda=6563$\AA) in emission, 
$H\delta(\lambda=4101$\AA) in absorption. The (*) symbol in Col.(5) indicates galaxies
for which $EW(H\alpha)$ and $S(H\alpha$) have been measured directly. Col.(6): $H\alpha$
  fluxes. Col.(7),(8),(9): $H\alpha$, 15-$\mu$m and 1.4-GHz luminosities. $H\alpha$ luminosities have been corrected for aperture losses but not
for extinction effects (see
  Section \ref{properti_sec} for more details). Col.(10): the spectral
  classification.
\end{minipage}
\end{table*}

The dominant class (16/22 $\sim$73\%) of the extragalactic sources is comprised
of galaxies characterized by star-formation at different levels. AGN (type 1
and 2) constitute $\sim$18\% (4/22) of the sample and early-type galaxies constitute
$\sim$9\% of the sample (only 2 objects have been found with no emission lines). In
the deeper fields (Lockman Hole, HDF-N: Fadda et al. 2002; Alexander et al.
2002; Elbaz et al. 2002) constraints on the AGN contribution to the mid-IR
sources have been provided from correlation analysis of deep X-ray and mid-IR
observations. Although the total fraction of AGNs (type 1 and 2) does not
appear to
change with decreasing infrared flux, the fraction of AGN1 with respect to
AGN2 (whose IR emission is probably dominated by star-formation activity in the host
galaxy) seems to decrease significantly at faint fluxes (although the numbers of
objects in all surveys is small). In fact, in S1 we
find that $\sim$15\% of identifications down to 1 mJy are AGN1, whereas in S2 (to
$\sim$0.7 mJy) this fraction is $\sim$9\% (2/22) and becomes still smaller in the deeper
Lockman field, $\sim$5\% to 0.25 mJy.  

In Figure \ref{spectra3_fig} we show the spectrum and the I and K$^{\prime}$ images of
source \#17. Contour levels of the 15-$\mu$m and of the radio emission are
plotted superimposed to the two images respectively. The source, identified
with a barred spiral at $z=0.02$, is shown separately because of the large
extension of its emission in the optical, radio and infrared bands (${\sim}0.5^{\prime}{\times}1^{\prime}$).
The infrared and radio emission are more pronounced in proximity of the spiral
arms with respect to the galaxy center. This is expected since IR and radio
emission in spiral galaxies are tracers of star-formation, which takes place
preferentially in the spiral arms. For this reason the spectrum, which is dominated by the galactic bulge, does not show any emission lines. To retain
consistent classification, this galaxy has
been classified as an early-type galaxy in Table \ref{spectra_tab}, despite its clear spiral morphology.

\section{Discussion}

\label{properti_sec}

\begin{figure}
\psfig{figure=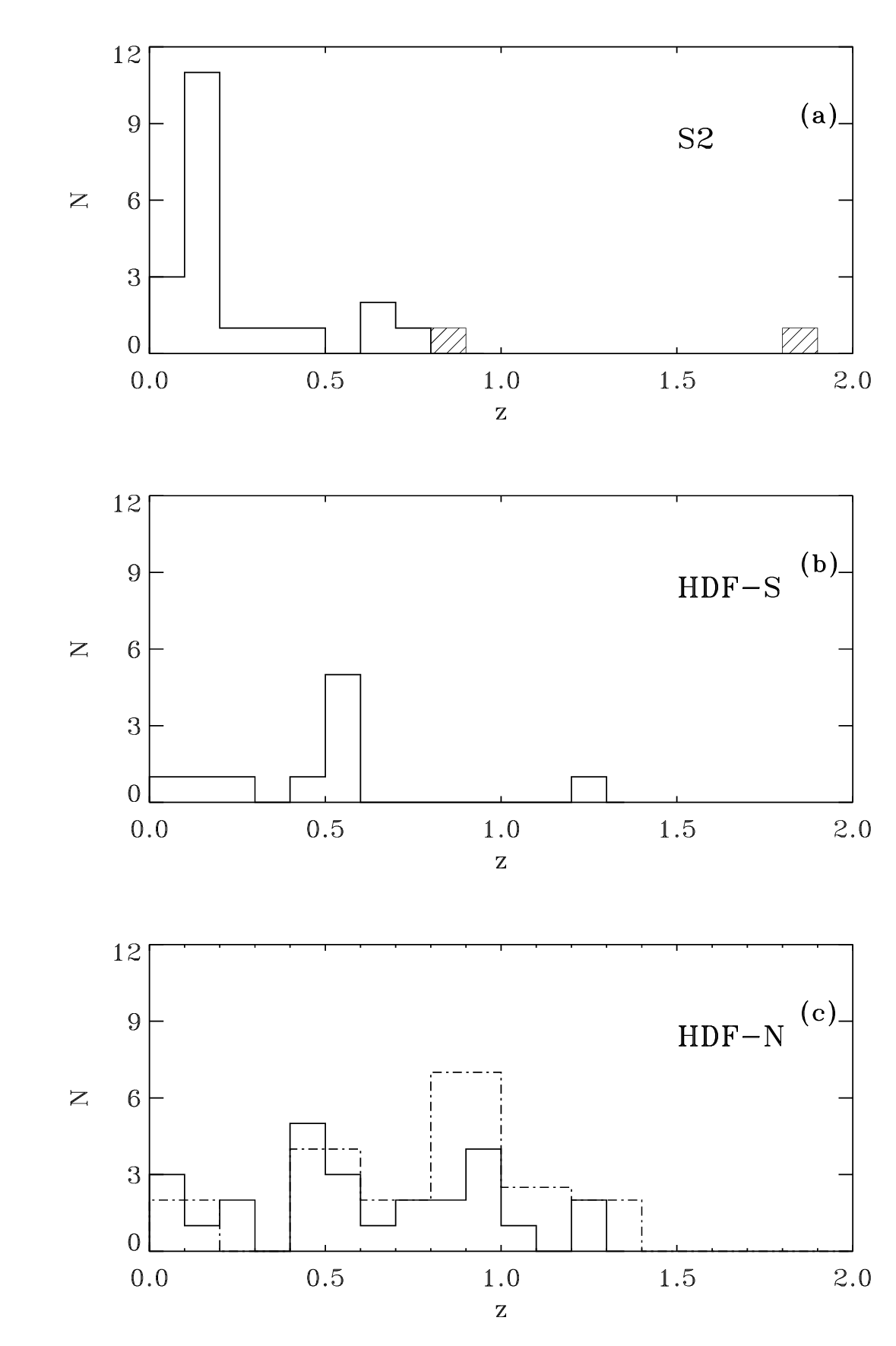,width=8cm}
\caption{Redshift distributions of the 15-$\mu$m sources in S2 ((a) panel, the hatched area represent the AGN1 objects), in the HDF-S ((b) panel, Mann et al. 2002) and in the HDF-N
  surveys ((c) panel, continuous line from Aussel et al. 1999; dashed line
  from Franceschini et al. 2001).}
\label{z_fig}
\end{figure}
\begin{figure*}
\centerline{
\psfig{figure=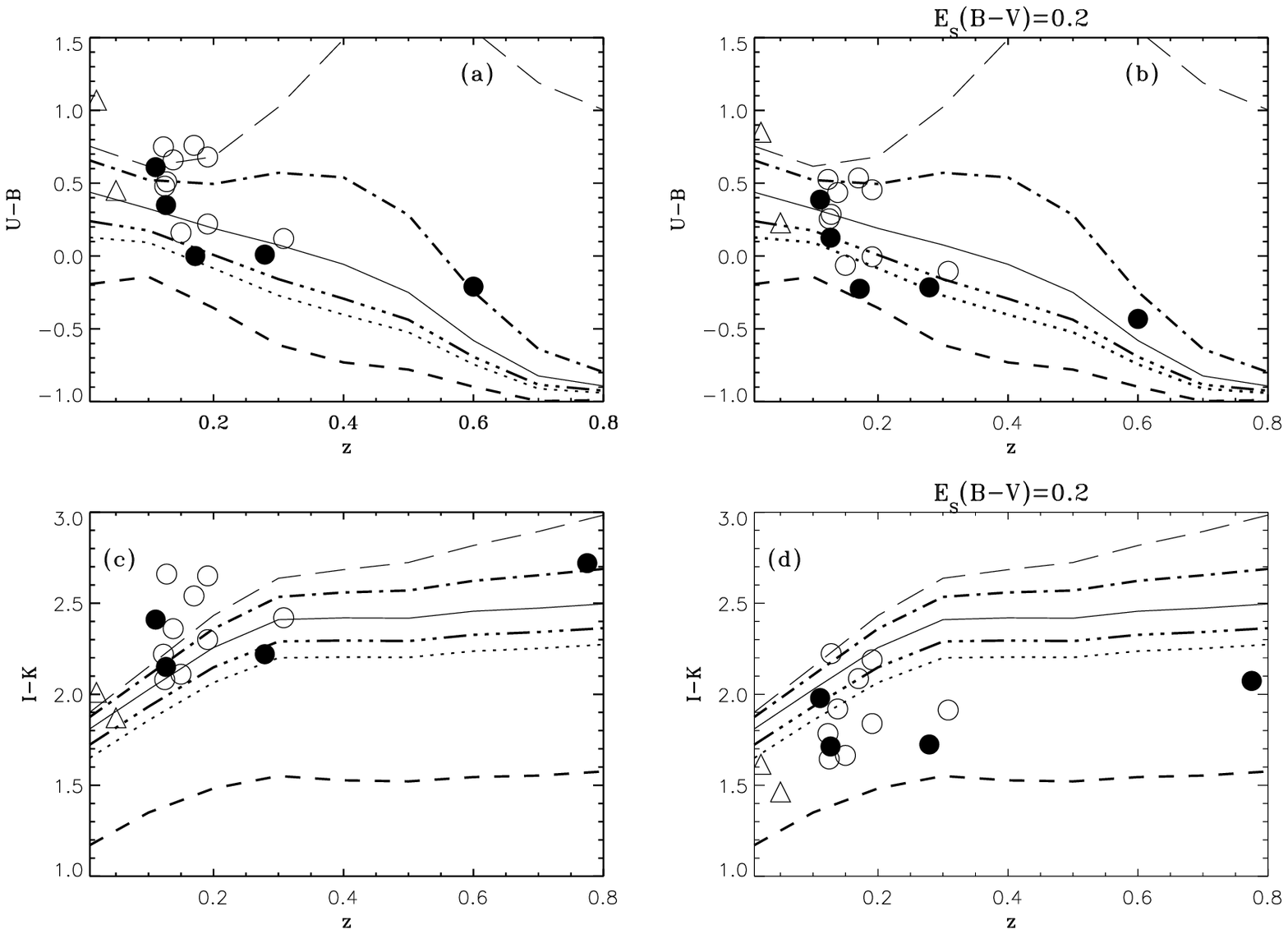,width=14cm}}
\caption{U-B ((a) and (b) panels) and I-K ((c) and (d) panels) colours vs. redshift diagrams for our galaxies. The different curves correspond to the colour-redshift
relations for evolved galaxies derived from Bruzual \& Charlot (1993) models and represent models for early-type galaxies (long-dashed line), S0 galaxies (dot-dashed
line),  Sab-Sbc spirals (solid and dot-dot-dot-dashed line), Scd (dotted line)
and Irregular (dashed line)
galaxies. The curves, kindly provided by
L. Pozzetti, have been computed using the appropriate filter (U,B, I and K) transmission used in our observations. The parameters of the models are given in Pozzetti et
al. (1996). In panels (a) and (c) no extinction correction has been
applied. In panels (b) and (d) colours have been corrected using the
reddening curve of Calzetti et al. (2000) with an extinction of $E_{s}(B-V)=0.2$.}
\label{colori_fig}
\end{figure*}

In this section we will describe the properties of our sample and, by making
comparisons with other IR surveys, we will highlight the contribution of this
work to the understanding of the nature of the IR sources.\\
First, the $z$ and 15-$\mu$m luminosity distributions of the S2 sources will be compared with those of deeper 15-$\mu$m fields, to look for evolutionary effects. Second, the extinction affecting the sources will be estimated and the results compared with those obtained from a sample of local high-luminosity IRAS galaxies, to study possible variation of the amount of dust with IR luminosity. In this context, we will discuss the possible dependence of reddening on IR luminosity by studying the relation between H$\alpha$ and IR luminosities and comparing our results with those found in a recent work by Kewley et al. (2002). Finally, we will compute and compare the star-formation rate of our galaxies derived from the IR, H$\alpha$ and radio indicators.

\subsection{Multi-band and spectral properties}
\label{multiband_sec}

In Figure \ref{mir_ir_fig} we present the I magnitude versus 15-$\mu$m flux and the 15-$\mu$m luminosity versus redshift diagrams. The objects are plotted with different
symbols according to their spectral classification, as described in the caption. Objects not observed spectroscopically are represented by filled triangles, while objects with no
optical counterpart to I $\sim$22 are represented by vertical arrows. Our
  data are compared
here with those from deeper surveys: HDF-S (crosses: Oliver et al. 2002, Mann
et al. 2002) and HDF-N (diagonal crosses: Aussel et al. 1999). In both
fields the percentage of spectroscopic identifications available in literature
is $\sim$50\%. All the galaxies of the HDF-S field have been identified in I
or near-IR bands (the objects not observed spectroscopically are represented
by crosses inside circles). For all the data, the 15-$\mu$m rest-frame luminosity have
been derived by assuming the Spectral Energy Distribution (SED) of M82 for
galaxies and AGN 2 and a typical Seyfert 1 SED for AGN 1 (Franceschini et
al. 2001). 

In the 15-$\mu$m$-$I diagram (Figure \ref{mir_ir_fig}, {\it Left panel}), the regions occupied respectively by stars and
by extragalactic objects are well separated by the dashed line which represents the MIR-to-optical ratio $R=2.5$, where $R$ is defined as
$R=S\times{10}^{\frac{m-12.5}{2.5}}$ ($S$ is the 15-$\mu$m flux in mJy while $m$ is the optical magnitude in the I band). As found also by Gruppioni et al. 2002, the
fraction of stars still keeps at $\sim$20-30
$\%$ for 15-$\mu$m flux densities fainter than $\sim$1 mJy. 

The starburst population is the dominant population at
faint 15-$\mu$m flux densities and weak optical magnitudes. The four S2 sources without I counterparts to I $\sim$22 and the two HDF-S 
   sources at fainter magnitudes (I $\sim$22), might belong to a separate population
   of objects characterized by faint optical magnitudes and/or high redshifts. A similar 
   indication is found also in the redshift-magnitude distribution of sources in the S1 survey 
   (La Franca et al. 2003, in preparation), where a second optically fainter
   population appears at R $\gsimeq$21.5, well separated from the bulk of the extragalactic 
   sample.

 In Figure \ref{z_fig} we
report the redshift distribution of the sources in the different surveys to
highlight the redshift ranges that these different samples cover.
 

As shown in Figure \ref{mir_ir_fig} ({\it Right panel})
 and Figure \ref{z_fig}, most of the spectroscopically identified
 extragalactic sources in S2 are at low-moderate
redshifts ($z\lsimeq$0.3), though 2 starburst galaxies and 2 AGN2 are up to
$z{\sim}$0.7. The higher redshift objects are, as expected, the two AGN1. In our analysis we consider AGN2 together with star-forming galaxies, according to the idea that for both
populations the IR spectrum may be dominated by starburst emission
(Franceschini et al. 2001). The median redshift of the S2 sources (excluding the 2 AGN1) is 0.17${\pm}$0.06
($\sigma_{med}=1.2533\frac{\sigma}{\sqrt{N}}$, Akin \& Colton 1970),
while in the HDF-S and HDF-N the median
redshifts significantly higher ($\sim$0.5 and $\sim$0.6 respectively).  

Considering the sample of sources with flux density
${\ge}0.8$ mJy (which is spectroscopically complete, see {\it Left} panel in
Figure \ref{mir_ir_fig}), we have compared the observed redshift distribution,
corrected for incompleteness by weighting each source for the corresponding
effective area, with the distribution predicted by the model fitting the
source counts (see Gruppioni et al. 2002). The model of Gruppioni et
al. (2002) has been obtained by re-adapting the Franceschini et al. (2001) model for which
the IR sources can be divided into three different populations with different
evolutionary properties: non-evolving normal spiral, strongly evolving
starburst plus AGN2, and evolving AGN1. While the observed and the predicted
$z$-distributions for the star-forming sources are in good agreement at $z\lsimeq$0.5 (first peak of the
predicted redshift distribution, see Figure 12 in Gruppioni et al. 2002), the
number of star-forming galaxies predicted by the model at $z$$>$0.5 it is higher ($\sim$50 \%) than observed ($\sim$20 \%). The discrepancy would be even larger without considering the luminosity break in the LLF of local
starburst galaxies proposed by Gruppioni et al. (2002). Disagreement between the predicted and the observed $z$-distributions is found
also in the HDF-S field, where the observed $z$ distribution (Mann et
al. 2002) is shifted to lower redshift than expected by the model and
  observed in the HDF-N (see Franceschini et al. 2001).
A possible cause for this disagreement between the observed and the predicted
distributions in both S2 and HDF-S could be the low statistics at high-$z$ in
S2 (only two sources with $S{\ge}0.8$ mJy have $z\gsimeq$0.5) or the spectroscopic incompleteness and the cosmic variance affecting small fields in the HDF-S (area $\sim$20 sq. arcmin). 
Only complete $z$-distributions in larger 15-$\mu$m fields (as ELAIS main fields or Lockman Shallow) not affected by small area variations will provide more stringent constraints for testing the model predictions. Only with such data it will be possible to eventually modify the model according to the observational constraints.
\begin{figure}
\psfig{figure=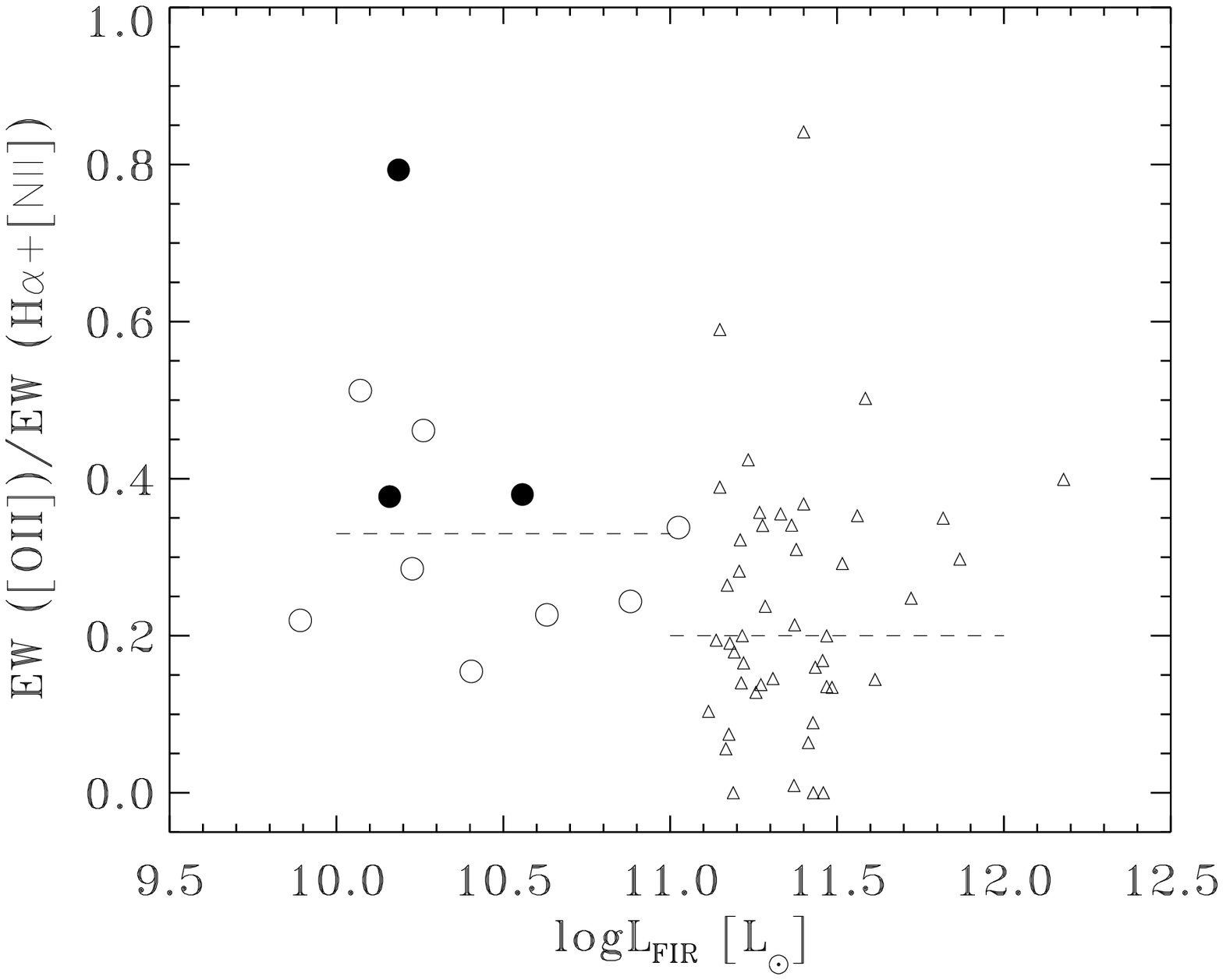,width=8cm}
\caption{$EW([OII])/EW(H\alpha+[NII])$ ratio as function of $L_{FIR}$ for our data (circles) and for the sample of
PW00 (empty triangles). The dashed lines correspond to the median
values of the ratio (0.33 and 0.21 respectively) for the two samples.}
\label{ew_o2_ha_fig}
\end{figure}
\begin{figure}
\psfig{figure=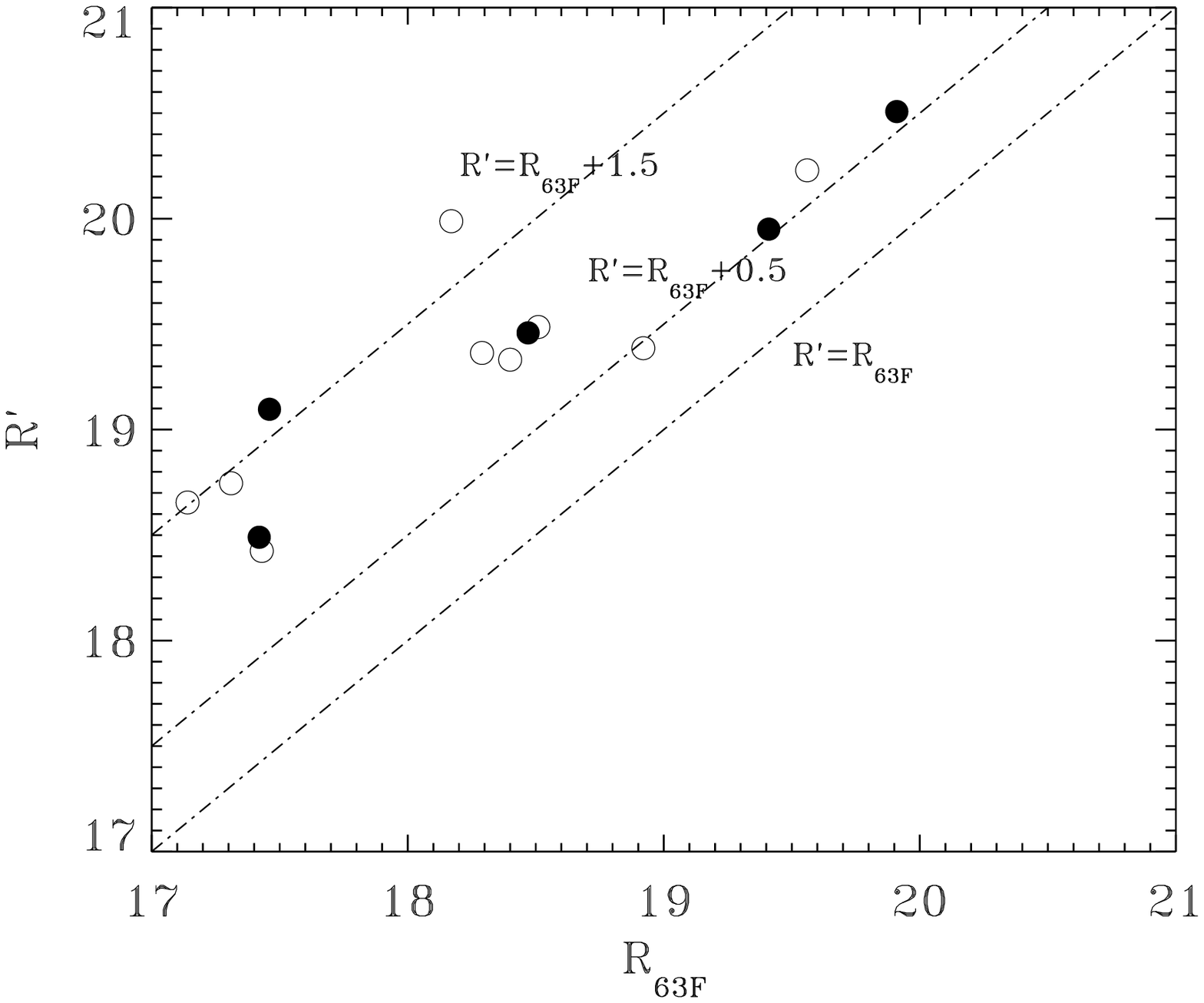,width=8cm}
\psfig{figure=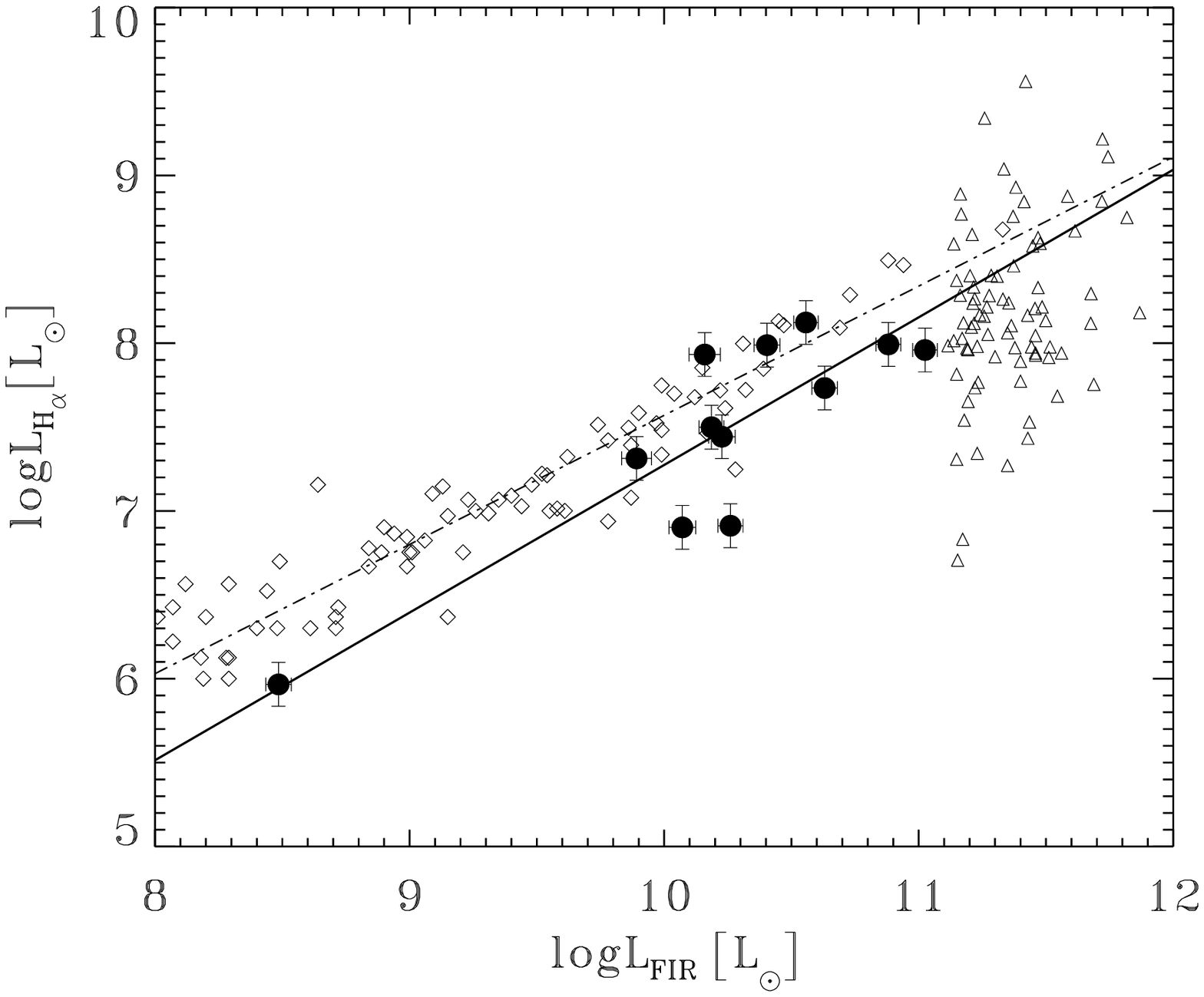,width=8cm}
\caption{{\it Top}: Relation between nuclear R magnitude (R$^\prime$) and total
R magnitude (R$_{63F}$). The three dot-dashed lines indicate differences of 0.0 (lower),
0.5 (central) and 1.5 mag (upper). {\it Bottom}: $H\alpha$ vs. FIR rest-frame luminosity (in
solar units). The filled circles represent our data with no
distinction between the different classes (AGN 1 have been excluded). The
empty diamonds are from Kewley et al. (2002) for a sample of galaxies detected by IRAS in the Nearby
Field Galaxy Survey (NFGS). The empty triangles are from PW00, after applying the larger possible aperture correction (a factor of 7). The solid and dot-dashed lines show respectively the fits between the two luminosities from this work and from Kewley et al. (2002).}
\label{lir_lha_fig}
\end{figure}

The majority of our galaxies + AGN2 objects (18/20,  $\sim$90 \%) have an
  IR luminosity ($L_{IR}=L[8-1000 \mu$m]) typical of ``starburst''
galaxies ($10^{10}L_{\odot}$$<$$L_{IR}{\le}10^{11}L_{\odot}$) or Luminous Infrared Galaxies (LIGs:
$10^{11}L_{\odot}$$<$$L_{IR}{\le}10^{12}L_{\odot}$), with no object with a
luminosity clearly in the range of those typical of Ultra Luminous Infrared Galaxies
(ULIGs: $L_{IR}$$>$$10^{12}L_{\odot}$). The IR luminosities have been
calculated from the 15-$\mu$m ones using the relation
$L_{IR}=11.1\times{L_{15{\mu}m}^{0.998}}$ (Elbaz et al. 2002). The median luminosity of our galaxy +
AGN2 sample is
 $L_{IR}=10^{10.8\pm{0.2}}L_{\odot}$. As shown in Figure \ref{mir_ir_fig} ({\it Right}) a similar range in luminosities is sampled also by
the deeper surveys (HDF-N and HDF-S), although for these surveys the bulk of
the distribution is shifted towards higher luminosities
(${\sim}10^{11.5}L_{\odot}$ in the HDF-N). Given the spectroscopic
  incompleteness of all the samples considered, the unidentified objects could be either
objects at similar redshift but absorbed or ULIGs at higher redshift. This suggests a possible trend of
luminosity with $z$, with the galaxies at
higher redshift being characterized by higher 15-$\mu$m luminosity. Such a
trend would be consistent with the sharp steepening observed in the 15-$\mu$m source counts
around 1-2 mJy (Elbaz et al. 1999; Gruppioni et al. 2002), which can be explained only under the
hypothesis of strong evolution (both in density and luminosity) for IR
galaxies. 

In Figure \ref{colori_fig}, the U-B and I-K colours vs. redshift plots are
shown. Panels (a) and (c) have no extinction correction,
while panels (b) and (d) have been obtained assuming an average extinction from
stellar continuum of $E_{s}(B-V)\sim$0.2. As shown in the plots, galaxies with
star-formation cover a wide range of colours but, if not corrected for
extinction, they are on average redder than
expected (a large fraction of objects are above the evolutionary curve of early-type
galaxies). By using the extinction curves of Calzetti et al. (2000), we have derived the extinction expected in different
bands for different values of the colour excess $E_{s}(B-V)$. This was done for each galaxy
separately, depending on its own redshift. An extinction of
$E_{s}(B-V)\sim$0.2$-$0.3 seems to be necessary to bring back the objects to
the intrinsic colours expected for late-type/starburst objects. Following Calzetti et
al. (2000), an extinction of $E_{s}(B-V)\sim$0.2$-$0.3 derived from the stellar
continuum corresponds to an extinction derived from the Balmer decrement
$H\alpha$/$H\beta$ of $E(B-V)\sim$0.4$-$0.6. Such amount of extinction is
lower than that found by PW00 analysing a sample of local Very Luminous
Infrared Galaxies (VLIRG: $L_{IR}>10^{11.15}L_{\odot}$, H$_0$=75 km s$^{-1}$
Mpc$^{-1}$ ), who derive $E(B-V)\sim$0.8$-$1.0 from the Balmer decrement $H\alpha$/$H\beta$. 
The indication that the extinction
affecting the lines is an increasing function of the IR luminosity is
highlighted also in Figure \ref{ew_o2_ha_fig}, where the $EW([OII])/EW(H\alpha+[NII])$ ratio
 is plotted as a function of $L_{FIR}$ for both our data and PW00 data (see
 next Section for the derivation of $L_{FIR}$ from $L_{{15\mu}m}$). The median values of the ratio are 0.33 and 0.21 respectively for S2 and
 VLIRG sample (the two distributions are different at a 2$\sigma$ level
from a K-S test). In the scenario proposed
by Poggianti et al. (1999) low values of the EW ratios are due to high dust
extinction, since the $[OII]$ emission line is more affected by dust than the
$H\alpha$ one. The continuum underlying the lines is produced by older,
less extincted stars and, by consequence, the net result is a low EW ratio of
the two lines. The anti-correlation between the $EW([OII])/EW(H\alpha+[NII])$
ratio and the IR luminosity (see also Kewley et al. 2002) is
consistent with the result found recently in the B band, where a similar trend between this ratio and the luminosity in the B band has been found (Jansen et
al. 2001, Charlot et al. 2002).

\subsection{Star-formation rates}

In this section we compute and compare the star-formation rate derived from three different
indicators: the FIR luminosity, the radio
luminosity and the $H\alpha$  luminosity.
The FIR luminosity has been derived by assuming a value of 5 for the
$L_{FIR}/L_{15{\mu}m}$ ratio. Such a ratio is an average value estimated in the ELAIS field S1,
representing typical starburst galaxies slightly more active than M82 (see also Elbaz et al. 2002). The radio
luminosity has been derived assuming a power-law spectrum with a
spectral index $\alpha\sim$0.7 ($S_{\nu}{\propto}{\nu}^{-\alpha}$). Finally, the
$H\alpha$ luminosity has been derived from the observed fluxes of the
$H\alpha$ line (see Table \ref{spectra_tab}). 

To correct for the flux losses
caused by observing only the nuclear region of the galaxies (the slit width is
1.0-1.5$^{\prime\prime}$, see Section \ref{spettro_sec}), we have assumed that the
$H\alpha$ line emission and the optical light are correlated over the entire
galaxy; we have then corrected the spectra fluxes for the ratio between the
R$_{63F}$ magnitudes from APM and the equivalent magnitudes derived from spectra (hereafter called R$^{\prime}$). We have estimated
the R$^{\prime}$ magnitudes by integrating the spectra over a standard Cousins R$_c$
filter and then we have reported the R$_c$ magnitude to APM magnitudes
(R$_{63F}$) following Bessel (1986)  (R$_c$-R$_{63F}$$\sim$0.1 for
R-I$\sim$0.7 as found for our
galaxies, see Table \ref{followup_table}). In Figure \ref{lir_lha_fig} ({\it Top}) the result
of the comparison is reported. The flux losses range from $\sim$0.5 mag for the weaker
sources (R$_{63F}$$\sim$19-20) to $\sim$1.5 mag for the brighter ones
(R$_{63F}$$\sim$17-18). This trend corresponds to a trend in the typical
angular sizes of the sources, from 1-2 arcsec (i.e. sources 8 and 13, Figure
\ref{spectra_fig}) to several arcsecs (i.e. sources 21 and 36, Figure
\ref{spectra_fig}). 
 Considering the sources with no APM counterpart, a correction of 0.5 mag has been applied to sources below the APM limit, while a correction of 1 mag has been applied to extended sources with unreliable APM magnitudes (see Table \ref{followup_table}).

\begin{figure*}
\centerline{
\psfig{figure=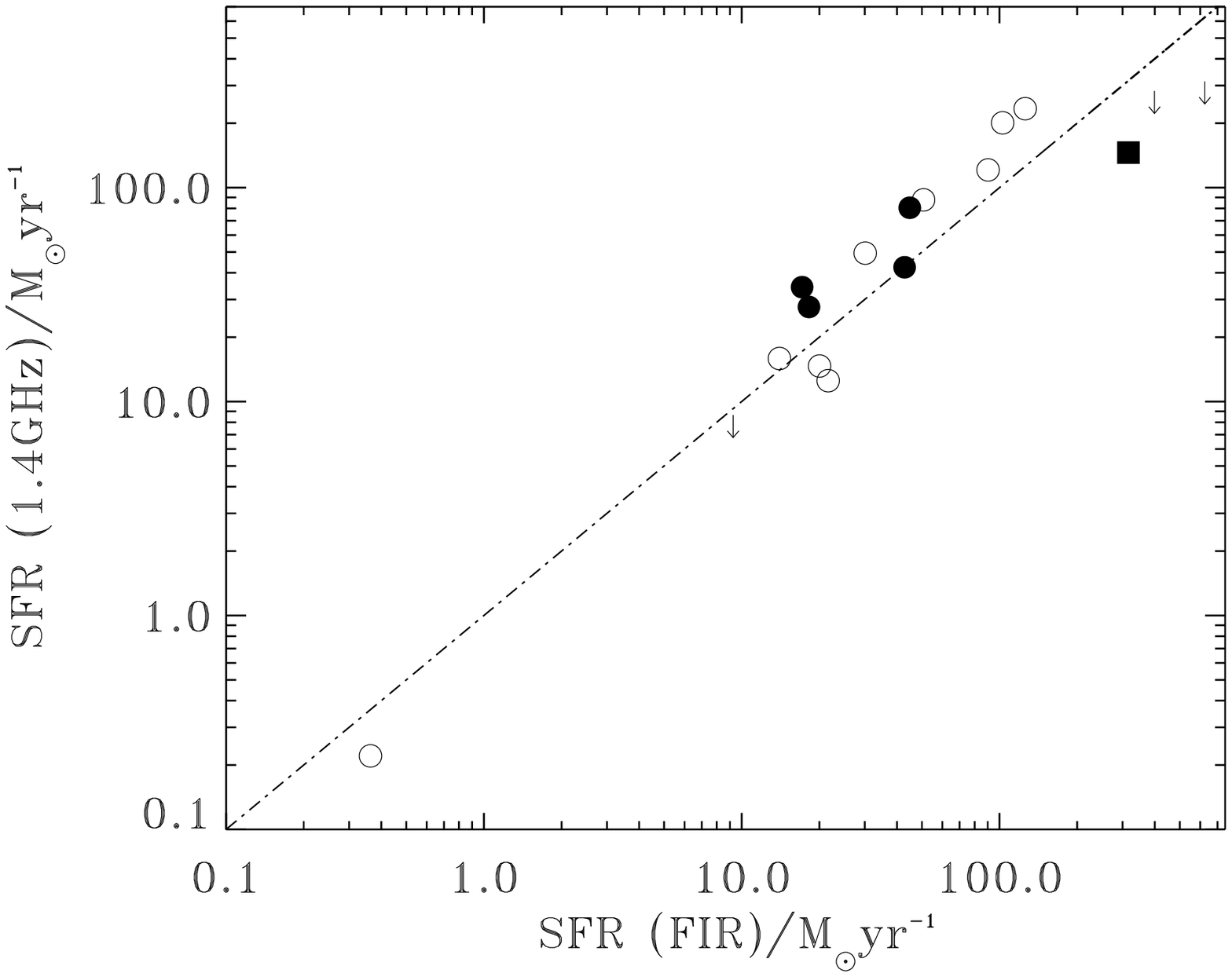,width=8cm}
\psfig{figure=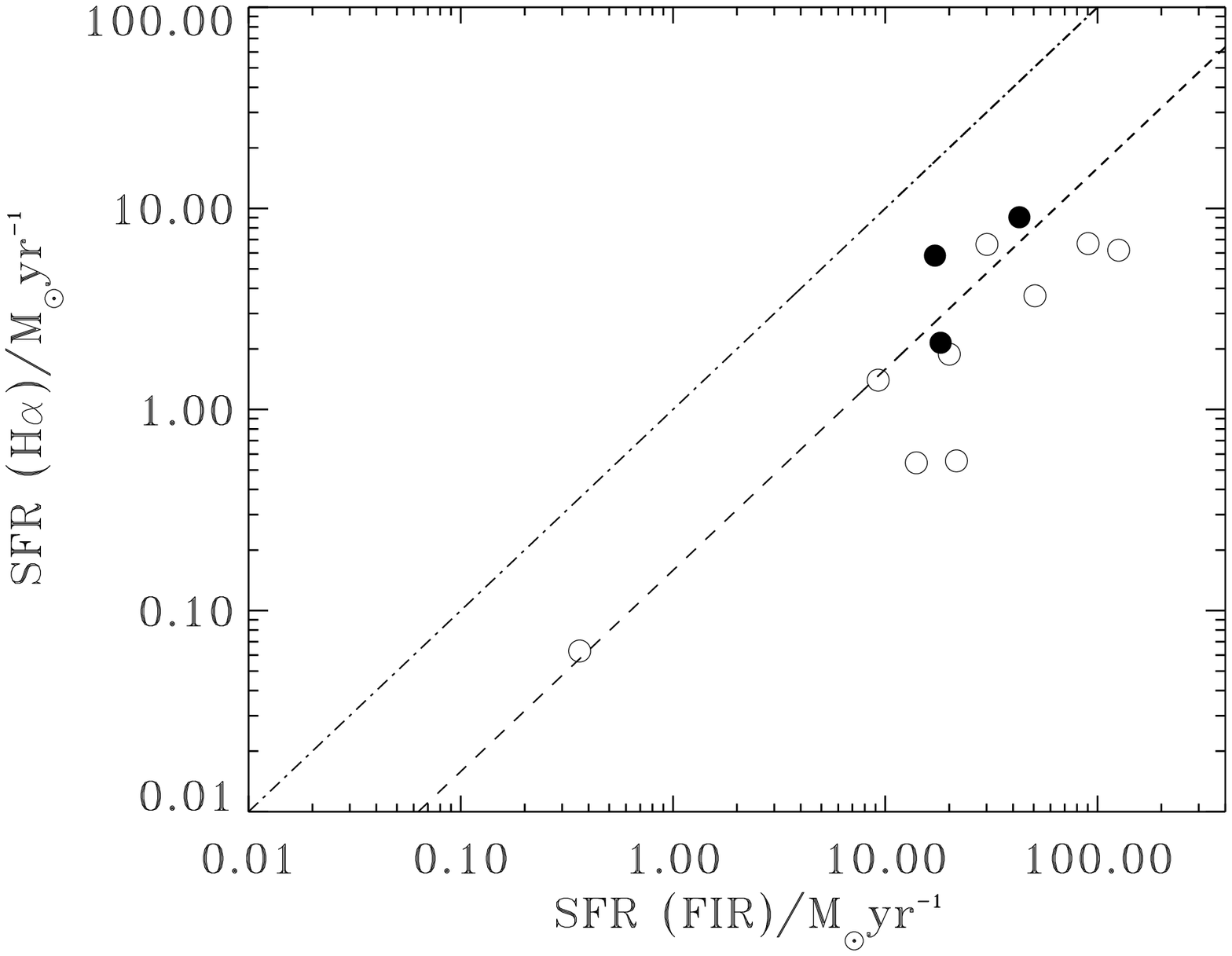,width=8cm}}
\caption{{\it Left:} SFR from 1.4 GHz vs. SFR from FIR. Symbols are the same
as in Figure \ref{mir_ir_fig}. The vertical arrows are the
radio 3$\sigma$ upper limits. The relation corresponding to
SFR(1.4GHz)=SFR(FIR) is shown as dot-dashed line. {\it Right:} SFR from
no-extinction corrected $H\alpha$ vs. SFR from FIR. The dot-dashed line shows
the relation SFR($H\alpha$)=SFR(FIR) and the dashed-line is found for an
average extinction of $E(B-V)=0.5$ [$A(H\alpha)=2$]. }
\label{sfr_fig}
\end{figure*}

Figure \ref{lir_lha_fig} ({\it Bottom}) shows the $H\alpha$ luminosity corrected
for aperture effects versus the FIR
luminosity. The filled circles represent our data with no
distinction between the different galaxy classes (type 1 AGN have been excluded). The
empty diamonds are the data of Kewley et al. (2002) for a sample of galaxies in the Nearby
Field Galaxy Survey (NFGS) with IRAS detections. The empty triangles are the data from PW00 after applying the maximum correction suggested by the authors (a factor of 7, see PW00). 
The plot shows a strong correlation between the two
luminosities over more than 4 orders of magnitudes; the large dispersion of the PW00 data is probably caused by the application of a unique average aperture
correction. The solid and dot-dashed lines show the best-fitting relations between the two
luminosities found respectively in this work and by Kewley et al. (2002) (in both cases using the {\it fitxy} routine from Numerical Recipes). This routine uses the linear least-squares minimization and includes error estimates for both variables. We derived the FIR luminosity uncertainties by propagating the measured errors on 15-$\mu$m fluxes (see Table \ref{catalogue_tab}), while we assumed errors of $\sim$30 \% for the $H\alpha$ luminosities (as estimated from the $EW$ measures). 
The resulting relation for our data has the form $log[L(H\alpha)]=(0.88\pm{0.07})log[L(FIR)]-(1.5{\pm}0.7)$. Though only indicative, our result shows a slope lower than 1 (2$\sigma$ level), in agreement with the result of Kewley et al. (2002) who find a slope equal to $0.77\pm0.04$. This suggests that the ratio between IR and $H\alpha$ luminosities is a
function of luminosity itself. Kewley et al. (2002) have shown how this effect could be
attributed to the reddening $E(B-V)$, finding that the IR to
$H\alpha$ luminosity ratio is a strong function of reddening $E(B-V)$ (see also Hopkins et al. 2001).

To convert the $H\alpha$ luminosities to star-formation rates, 
we use the calibrations given by Kennicutt (1998) for a Salpeter (1955) initial
mass function (IMF) running from 0.1 to 100 M$_{\odot}$. To convert the
IR and 1.4-GHz luminosities to star-formation rates, we use the
calibration given by Cram et al. (1998) (see also Condon 1992), scaled to the same IMF and mass range of
Kennicutt (1998) following Mann et al. (2002).\footnote{Condon (1992) relations are based on
a Miller \& Scalo (1979) IMF over [5,100] M$_{\odot}$.} 

The set of estimators considered are:

\begin{displaymath}
SFR_{H\alpha}=7.9{\times}10^{-35}L(H\alpha)(W)
\end{displaymath}

\begin{displaymath}
SFR_{FIR}=0.9{\times}10^{-23}L_{\nu}(60{\mu}m)(WHz^{-1})
\end{displaymath}

\begin{displaymath}
SFR_{1.4GHz}=1.1{\times}10^{-21}({\nu}(GHz))^{\alpha}L_{\nu}(1.4GHz)(WHz^{-1})
\end{displaymath}
The FIR luminosity has been converted to 60-$\mu$m luminosity
($L_{FIR}\sim$$1.3{\times}L_{60{\mu}m}$), by considering the definition of the FIR flux given by Helou et al. (1998) ($FIR=1.26[2.58S_{60{\mu}m}+S_{100{\mu}m}]10^{-14}$ [Wm$^{-2}$]), the FIR$-$1.4-GHz relation (`q'
parameter; Condon 1992, Yun et al. 2001) and the 1.4-GHz$-$60 $\mu$m
relation ($S_{60{\mu}m}\sim$$127S_{1.4GHz}$, Cram et al. 1998).

 In Figure \ref{sfr_fig}, the SFR derived from 1.4 GHz ({\it Left}) and
from $H\alpha$ ({\it Right}) are
compared to the SFR obtained from FIR emission. The $H\alpha$ estimates are
not corrected for extinction. As found by Condon (1992), the
star-formation rates deduced from the FIR and the decimetric radio luminosities are well correlated. Moreover, approximately the same
relation found locally by Condon (1992) is still valid at the redshifts
sampled by our data ($z_{med}\sim0.2$). For a detailed study of the
radio/Mid-IR correlation see the study of Gruppioni et al. (2003) for
the ELAIS sources and the works of Garrett et al. (2002) and Bauer et al.
(2002) for the HDF-N and the CDF-N sources.
 \\

The same level of agreement is not found between the $H\alpha$ and the FIR
indicators, since the SFR based on $H\alpha$ are a factor of $\sim$5-10 lower
than the SFR based on FIR (see also Cram et al. 1998 for a detailed
discussion). The two indicators agree after applying the mean extinction
correction suggested by the Balmer decrements and the optical colours
($E(B-V)\sim$0.5, $A(H\alpha)\sim$2.0) to the $H\alpha$ estimates. A correction of $\sim$2 mag is intermediate between the correction of 1 mag found by Kennicutt (1983) for a sample of local spirals and the correction of $\sim$2.5-3 mag found by
PW00 for the sample of local Very Luminous Infrared Galaxies, indicating, as said
before, that the extinction suffered by galaxies is an increasing function of
IR luminosity. 

A further evidence supporting this hypothesis is supplied by the comparison of the SFR
rates computed in this work with those of Rigopoulou et al. (2000), who have
studied the infrared and $H\alpha$ properties of a subsample of HDF-S ISOCAM
galaxies at high redshift and high IR luminosities ($<z>{\sim}0.62$,
$<L_{FIR}>{\sim}10^{11.4}$, H$_0$=75 km s$^{-1}$ Mpc$^{-1}$). They find that SFR estimates
from not-corrected $H\alpha$ are lower than the FIR ones by a factor of 5-50 and that a reddening correction of $\sim$3
mag is necessary to bring the $H\alpha$ and FIR rates to the same scale.

\section{Conclusions}

We have studied the properties of a new sample of 15-$\mu$m sources detected
in the ELAIS Southern field, S2, using the {\it Lari method} of reduction (see
L01). The 15-$\mu$m catalogue is constituted by 43 sources with fluxes between 0.4
and 10 mJy and hence is perfectly suited to investigate the properties of
sources in the flux density region where the 15-$\mu$m source
counts are observed to diverge from no-Euclidean predictions (2-3 mJy: Elbaz et al. 1999,
Gruppioni et al. 2002). 

A high percentage of sources ($\sim$90\%) have a reliable optical identification brighter than
$I\sim$21 mag and for 21 objects a radio counterpart has been found. Eight
bright sources have been classified as stars from the photometric data, while spectra have
been obtained for 22 extragalactic objects, reaching a high identification
percentage (30/43, $\sim$70\%). All but one of the 28 sources
with 15-$\mu$m flux density $>$ 0.7 mJy are identified. Given the good spectral quality,
we were able to determine object type, redshift, equivalent width of the main
lines and $H\alpha$ fluxes. The majority of the sources are star-forming
galaxies (18 out of 22 objects, $\sim$82 \%) with 15-$\mu$m luminosities typical of ``starburst''
and Luminous Infrared Galaxies (LIGs). AGN (type 1 and 2) constitute $\sim$18\% of the
extragalactic sample, while the fraction of early-type galaxies is small
($\sim$9\%). The median luminosity of our galaxies + AGN2 objects is
 $L_{IR}=10^{10.8\pm{0.2}}L_{\odot}$ while their median redshift is
  0.17${\pm}$0.06.

In comparison with deeper published 15-$\mu$m surveys (HDF-S: Oliver et al. 2002, Mann et al.
2002; HDF-N: Aussel et al. 1999), the S2 galaxies are at a lower redshift ($z_{med}\sim$0.2
 instead of 0.5, and 0.6) and lower luminosities (the HDF-N has a peak
around ${\sim}10^{11.5}L_{\odot}$). However, a more detailed comparison of the
properties of the sources in the three samples would require a higher level of
completeness in the spectroscopic identification (S2 is $\sim$70\% complete, while HDF-S, HDF-N are
 $\sim$50\% complete).

We have compared the 15-$\mu$m and $H\alpha$ luminosities corrected for
aperture losses but not for extinction. Consistently with the results from Kewley et al. (2002), we have
found an indication that the ratio between 15-$\mu$m and $H\alpha$ luminosities not
corrected for extinction increases with luminosity, suggesting that the amount of reddening is
not constant, but is a function of luminosity, the more luminous galaxies being the more
affected by extinction.

Using the 15-$\mu$m, radio and $H\alpha$ luminosities, we have derived the
star-formation rates and we have compared
the results obtained with the three different indicators. We have found a tight linear relation between radio and 15-$\mu$m based
star-formation estimates. This indicates that the well studied local radio/FIR correlation (Condon 1992) is still valid at the median redshift of our sample
($z_{med}\sim$0.2). The derived star-formation values typically range between 10 and 200 M$_{\odot}$yr$^{-1}$, with a median value of $\sim$40 M$_{\odot}$yr$^{-1}$. Estimates based on the H$\alpha$ luminosities tend to be more
scattered and to lie about a factor of 10 below the trend defined
by the radio/FIR estimates. An
average correction of $\sim$2 mag is necessary to bring the $H\alpha$ estimates on the same
scale of the other indicators. This correction is consistent with the
 average extinction correction derived from the Balmer decrements and from the optical colours (assuming the
Calzetti et al. 2000 reddening curve) and is intermediate between the extinction found for local spiral galaxies (Kennicutt
1983) and the extinction found by PW00 for a sample of local very luminous infrared galaxies.

\section*{Acknowledgments}
This research was supported by the Italian Ministry for University and
Research (MURST) under grants COFIN01 and by the Agenzia Spaziale Italiana
(ASI) under the contract ASI-I/R/27/00. We thank Lucia Pozzetti for kindly
providing colours-redshift models for different type of galaxies and helpful
discussion, and Lisa Kewley for kindly providing data relative to the Nearby
Field Galaxy Survey (NFGS). We thank the referee David Alexander for his careful reading
of the manuscript, interesting comments and suggestions which greatly improved
the quality of the paper.

\end{document}